\documentclass[acmsmall]{acmart}

\usepackage{algorithmic}
\usepackage{graphicx}
\usepackage{textcomp}
\usepackage{paralist}
\usepackage{xcolor}

\usepackage[utf8]{inputenc}

\usepackage{mathtools}
\usepackage{csquotes}
\usepackage[inline]{enumitem}
\usepackage{nicefrac}
\usepackage[capitalize,noabbrev]{cleveref}
\usepackage{bm}
\usepackage{amsthm}

\usepackage{xcolor}

\newcommand{\margin}{\epsilon} 
\newcommand{\smallprob}{\delta} 

\DeclareMathOperator*\Mean{\mathrm{Mean}}
\DeclareMathOperator*\LMean{\mathrm{LMean}}
\newcommand{\supl}{\mathrm{Sup}}

\newcommand\Aggo[1]{\mathrm{Agg}{[}#1]}
\newcommand\val{\mathrm{val}}

\newcommand\slope{L}
\newcommand\featdim{D}

\newcommand\Ext{\mathsf{Ext}}

\newcommand{\FeatSp}{\ensuremath{\mathsf{FeatSp}}}

\newcommand{\R}{\ensuremath{\mathbb{R}}}
\newcommand{\reals}{\R}

\newcommand{\N}{\ensuremath{\mathbb{N}}}

\newcommand{\concat}[2]{#1#2}
\newcommand{\tuplevec}[1]{\bar{\bm{#1}}}
\DeclareMathOperator{\ball}{\mathrm{B}}

\newtheorem{theorem}{Theorem}
\newtheorem{fact}[theorem]{Fact}
\newtheorem{obs}[theorem]{Observation}

\newcommand\GrTp{\mathrm{GrTp}}

\newcommand\GrTpSp{\mathsf{GrTp}}

\DeclareMathOperator\Ne{\mathcal N}

\newcommand{\coreof}[2]{#1\vert_{#2}}

\let\Pr\undefined
\DeclareMathOperator*\Pr{\mathbb{P}}
\DeclareMathOperator*\Ex{\mathbb{E}}

\newcommand{\wt}{\widetilde}

\newcommand{\lam}{\lambda}

\newcommand\lra\leftrightarrow
\newcommand\Lra\Leftrightarrow
\newcommand\dsb\doublesqbracket

\DeclarePairedDelimiter{\doublesqbracket}{\llbracket}{\rrbracket}
\DeclarePairedDelimiter{\abs}{|}{|}
\DeclarePairedDelimiter{\norm}{\lVert}{\rVert}

\def\erdosrenyi{Erd\"os-R\'enyi}

\newcommand{\Gcal}{\mathcal{G}}
\newcommand{\Tcal}{\mathcal{T}}

\newcommand{\Pcal}{\mathcal{P}}
\newcommand{\Qcal}{\mathcal{Q}}
\newcommand{\BP}{\mathrm{BP}}
\newcommand{\FBP}{\mathrm{FBP}}
\newcommand{\Dcal}{\mathcal{D}}

\newcommand{\srank}{\mathrm{Srank}}
\newcommand{\mrank}{\mathrm{Mrank}}
\newcommand{\lmrank}{\mathrm{LMrank}}
\newcommand{\rank}{\mathrm{rank}}

\newcommand{\dist}{{}}
\newcommand{\Po}{\mathrm{Po}}

\DeclareMathOperator{\Bin}{Bin}

\newcommand{\NN}{\mathbb{N}}
\newcommand{\GG}{\mathbb{G}}
\newcommand{\HH}{\mathbb{H}}
\newcommand{\TT}{\mathbb{T}}
\begin{document}
\title{Convergence Laws for Extensions of First-Order Logic with Averaging}
\author{Sam Adam-Day}
\email{me@samadamday.com}
\affiliation{%
  \institution{Oxford University Department of Computer Science}
  \city{Oxford}
  \country{United Kingdom}
  }
\author{Michael Benedikt}
\email{michael.benedikt@cs.ox.ac.uk}
\affiliation{%
  \institution{Oxford University Department of Computer Science}
  \city{Oxford}
  \country{United Kingdom}
}
\author{Alberto Larrauri}
\email{alberto.laurrari@cs.ox.ac.uk}
\affiliation{%
  \institution{Oxford University Department of Computer Science}
  \city{Oxford}
  \country{United Kingdom}
}
\date{}

\begin{abstract}
    For many standard models of random structure, first-order logic sentences exhibit  a convergence phenomenon on random inputs. The most well-known example is for random graphs with constant edge probability, where the probabilities of first-order sentences converge to 0 or 1. In other cases, such as certain ``sparse random graph'' models, the probabilities of sentences converge, although not necessarily to 0 or 1. In this work we deal with extensions of first-order logic with aggregate operators, variations of averaging. These logics will consist of real-valued terms, and we allow arbitrary Lipschitz functions to be used as ``connectives''. We show that some of the well-known convergence laws extend to this setting.
\end{abstract}

\maketitle
\section{Introduction} \label{sec:intro}

For many standard random graph models, first-order logic sentences exhibit convergence over random inputs.
The most well-known example is for the \erdosrenyi\ random graph model with constant edge probability, where probabilities converge asymptotically almost surely to zero or one \cite{faginzeroone, glebskii}: the ``zero-one law for first-order logic''.  Zero-one laws for first-order logic have been established both for other \erdosrenyi\ probabilities \cite{shelahspencersparse}, and for a uniform distribution over restricted structures \cite{comptonsurvey,kolaitisfreegraphs,zeroonelawmaps}. In several other settings --- for example, words with the uniform distribution, or \erdosrenyi\ graphs with linearly decaying edge probability --- we have a convergence law but not a zero-one law \cite{lynchdegreesequence, lynchlinearsparse}. In other words, the probability of each sentence of first-order logic convergences, but not necessarily to zero or one.

Beyond first-order logic, asymptotic behavior has been investigated for infinitary logic \cite{kolaitisvardiinfty}, for extensions with a parity test \cite{kolaitisparity}, and for fragments of second-order logic \cite{kolaitisvardiso}.
But, to the best of our knowledge,
the probabilistic behavior of first-order logic extended with \emph{aggregate operators}, like average, has been studied in a very limited capacity despite the fact that these play a key role in practical languages like SQL, as well as in graph learning models. Two exceptions to this statement are \cite{koponen, koponen2024relative}, following up on earlier work in \cite{jaeger,koponentcs}: we defer a discussion of these papers to the related work section.

In this paper we consider a real-valued logic that extends first-order logic with aggregates, in the same spirit as ``continuous logic'' \cite{clitay,continuouslogicck}. Like in continuous logic, we allow our input structures to contain real-valued functions: for simplicity, we stick to graphs in which nodes are annotated with unary real-valued functions. Each term in the logic defines a bounded real-valued function. We allow arbitrary Lipschitz functions as ``connectives'', which in particular allows us to capture Boolean operators. And we include supremum and infimum, which generalize existential and universal quantifiers to the real-valued setting. 

We fix distributions $P_n$ on input structures of each size $n$, generalizing the standard random graph model.  Our terms thus generate a sequence of random variables indexed by $n$, and we identify situations where this sequence converges using standard notions of random variable convergence.
Our results extend several classical convergence laws for first-order logic, while also extending recent results for real-valued logics that have aggregation as the \emph{only} quantification \cite{adamday2023zeroone,usneurips24}. Together with \cite{koponen,koponen2024relative}, these are the only convergence results we know of for proper extensions of first-order logic supporting aggregation. 

Our results show something stronger than convergence for closed terms: we prove that
for a broad term language with free variables, aggregate operators can be eliminated asymptotically almost surely. See Theorem \ref{thm:denseaggelim} in the case of random dense graphs, and Theorem \ref{thm:aggelimsparse} for
random sparse graphs.

\paragraph{Organization} After an overview of related work in Section \ref{sec:related} and preliminaries in Section \ref{sec:prelims}, we present our results on dense {\erdosrenyi} in Section \ref{sec:dense}, followed by the more involved sparse case in Section \ref{sec:linearsparse}.  We close with discussion in Section \ref{sec:conc}.

{\bf Acknowledgements}

The research leading to these results was supported by UKRI
EP/X024431/1.

This work is the full version of the extended abstract \cite{uslics}. We thank the reviewers of LICS 2025 for many comments that have been incorporated in the manuscript.
\section{Related work} \label{sec:related}
We mentioned that one starting point for our work are convergence laws for first-order logic, originating with \cite{faginzeroone, glebskii}. We review what is known for first-order logic for the standard {\erdosrenyi} graph model in Section \ref{sec:prelims}.
Two orthogonal extensions of logical convergence laws are: to other probability distributions on graphs,
and to other logics.

In terms of other distributions, Compton initiated the study of the uniform distribution over restricted graph classes, and this line has continued with a number of zero-one and convergence laws for first-order logic and monadic second-order logic on restricted structures, see, e.g.\@ \cite{kolaitisfreegraphs,noyminorclosed}. Recently asymptotics of first-order logic have been investigated for a model that is not uniform, but where the probability of an edge is not independent: so-called ``preferential attachment'' distributions \cite{attachment}. In terms of logics, fragments of monadic second-order logic have received extensive attention \cite{kolaitisvardiso}, while infinitary logic has been studied mainly for the classical {\erdosrenyi} setting \cite{kolaitisvardiinfty}.
Logics with a parity generalized quantifier are studied in \cite{kolaitisparity}. Recently there have been convergence results for continuous logic on metric spaces \cite{zeroonecl} and for semi-ring valued logics \cite{zereonelawsemiring}: these logics do not allow modeling the aggregation we deal with, i.e.\@ averaging over real-valued structures.

Logics with probability quantifiers -- where the models are equipped with a probability distribution, and the formulas are allowed to refer to these distributions,  are studied in \cite{almostsurehjk}. Convergence is proven only when certain ``critical values'' are avoided within the formulas. In a very different context, a similar restriction occurs in \cite{adamday2023zeroone}: there the convergence results are for graph neural networks (GNNs). The GNNs in the paper return Booleans, and the convergence results require that the decision boundaries between true and false avoid certain values.

We now discuss the most closely-related lines of work, originating with \cite{jaeger}, and including the recent \cite{ koponen,koponen2024relative}, which builds on the earlier \cite{koponentcs}. As mentioned in the introduction, these contain the only
other convergence results we know of for term languages supporting aggregation that extend first-order logic.
The main theorem of \cite{koponen} is an almost-sure aggregation elimination result for a term language called PLA, over a family of distributions on relational structures called \emph{lifted Bayesian Networks} (LBNs below). The aggregation elimination result applies to ``admisssible LBNs'', which can express a wide range of probability distributions over relational structures, and they do not require the PLA formulas to avoid critical values.
Crucially, admissible LGNs subsume Erd\"os-R\'enyi graphs with constant edge probability. But they do not capture sparse variants, like the linear sparse case that we consider in the second part of our paper. A language similar to PLA, but two-valued, was defined in \cite{jaeger}, and a convergence result was proved there for a very restricted family of LBNs. The PLA language of \cite{koponen} consists of terms that take values in $[0,1]$, built up with real-valued connectives extending the Boolean operators, and supporting a wide range of aggregation functions, including conditional mean and supremum. 
Because of incomparability at the level of distributions, our results do not subsume those of \cite{koponen}. The results of \cite{koponen} extend our theorem for the dense case (\Cref{thm:denseaggelim}), with two caveats. A minor caveat is that PLA does not support general Lipschitz functions as  we do. However, the proof in \cite{koponen} could be easily extended to accommodate this richer class  of functions. The second caveat is that our language allows real valued functions on input nodes, while \cite{koponen} does not. We imagine that the two results could easily be unified to support a variant of ``real-valued'' LBNs, although we have not investigated this.

We now turn to \cite{koponen2024relative}. This provides another aggregation elimination result (Theorem 5.11) for a very expressive family of probability distributions, which now subsumes sparse Erd\"os-R\'enyi graphs. But the logic involved does not extend first-order logic. Theorem 8.6 of the paper provides convergence results for a much richer logic, which can express Boolean statements about percentages. But now the formulas must be restricted to avoid using certain values, in the same spirit as the critical values of \cite{almostsurehjk} mentioned earlier.

As mentioned in the introduction, \cite{usneurips24} provides convergence results on a term language which does not subsume first-order logic, since it has \emph{only averaging} as the quantifier. This is in the same spirit as Theorem 5.11 of \cite{koponen2024relative} The distributions involved are much more general than those in this paper, including several where a convergence law for first-order logic is known to fail. The motivation in \cite{usneurips24} is to model flavors of GNN that return real values, and the paper includes an empirical study of the convergence rates on examples arising from GNNs.

\section{Preliminaries} \label{sec:prelims}

\paragraph{Conventions}
Given $n\in \NN$ we write $[n]$ for the set $\{1,\dots, n\}$. Over-lined variables,
such as $\bar v$, represent finite tuples of arbitrary length $|\bar v|$. 


\paragraph{Graphs and featured graphs}
A graph $G$ consists of a finite set of vertices $V(G)$ and a set of undirected edges $E(G) \subseteq \binom{V(G)}{2}$ with no loops.
The maximum degree of a vertex in $G$ is denoted by $\Delta(G)$. Given two vertices $u,v\in V(G)$, the distance between them $d_G(u,v)$ (or just
$d(u,v)$ when $G$ is clear from the context) is the minimum number of edges in a path connecting $u$ and $v$, or infinity if such a path does not exist. For any vertex $u$, its neighborhood is $\Ne(u) \coloneqq \{u \in V(G) \mid (u, v) \in E(G)\}$.

As mentioned in the introduction, since our logical languages allow us to manipulate numbers,  we also allow numerical data as part of the input graphs, as is common in real-world graphs. 
A \emph{multi-rooted featured graph}
(MRFG) $\GG$ is a tuple $(G,\bar v, \chi)$ where $G$ is a graph, $\bar v$ is a tuple of vertices in $G$ called \emph{roots}, and $\chi$ is a map from $V(G)$ to $\reals^\featdim$ for some $\featdim$. 
When $\bar v$ is the empty tuple we call $(G, \chi)$ a \emph{featured graph}.  Note that \emph{in this paper, all graphs, hence all MRFGs, will be finite}.
Given a vertex $u\in V(G)$, we denote by $\GG[u]$ the MRFG obtained by appending $u$ to $\GG$'s list of roots. That is, $\GG=(G,\bar v u, \chi)$.
The usual notions from graph theory are extended to MRFGs in the natural way.
\par
Given a vertex $v$ in a MRFG $\GG$ and a number $r\in \NN$ we write $\ball_r^{\GG}(v)$ (or just
$\ball_r(v)$ when $\GG$ is clear from the context) to denote the MFRG obtained by restricting the underlying graph of $\GG$ to the vertices that are at distance at most $r$ from $v$, considering $v$ as the only root, and restricting $\GG$'s feature function to the new set of vertices. \par

\paragraph{Random graphs and random featured graphs}
A \emph{random graph model} defines, for every number $n\in \NN$, a distribution over graphs with the set of vertices $[n]$.

A \emph{random featured graph model} is defined similarly, where for any $n$ we have a distribution over random featured graphs with $n$ vertices and $\featdim$ features, such that for any fixed $n$-vertex graph $G$ and any open set $S$ in $(\reals^\featdim)^n$, the set of random featured graphs extending $G$ with feature vector in $S$ is measurable. 
 Our random featured graph models will always be obtained by independently combining a random graph model and a distribution $\Dcal$ over the standard Borel sigma-algebra over $\R^\featdim$ with \emph{bounded support}: we refer to the latter as a \emph{feature distribution}. We define $\FeatSp \subseteq \reals^\featdim$ as the set of accumulation points of $\Dcal$'s support. In other words, $\FeatSp$ is the set of points $x$ such that every open ball centered at $x$ has non-zero probability according to $\Dcal$. Observe that $\FeatSp$ is a compact set.  \par


\paragraph{Probability theory background}
We say that a sequence of events $(A_n)_{n\in \NN}$ holds asymptotically almost surely (abbreviated to a.a.s.) if $\lim_{n\to \infty} \Pr(A_n)=1$. \par
Let $(X_n)_{n\in \NN}$, $(Y_n)_{n\in \NN}$ be two sequences of real-valued random variables over the same probability space. We say that $X_n$ \emph{converges in probability} to $Y_n$, denoted $X_n \xrightarrow{p} Y_n$ if
$\lim_{n\to \infty} \Pr(\abs*{X_n- Y_n} \geq \epsilon) = 0$
for any $\epsilon>0$. Let $Z$ be another real-valued random variable. We say that $X_n$ \emph{converges in distribution} to $Z$, denoted $X_n \xrightarrow{D} Z$, if 
for any real number $x$ that is a continuity point of $z\mapsto \Pr(Z \leq z)$
it holds that $\lim_{n\to \infty} \Pr(X_n \leq x) = \Pr(Z \leq x)$.
Convergence in probability is a stronger notion than convergence in distribution.





\paragraph{{\erdosrenyi} random graphs and featured graphs}
Our random graph models will be based on the standard \erdosrenyi\ distribution.
For $p$ a function from natural numbers to $[0,1]$, the \erdosrenyi\ distribution $\Gcal(n,p)$ is a random graph model defined as follows: for a given $n$,  the graph can be taken to have vertex set $[n]$, and for each distinct $i, j \in [n]$, we have that $i$ and $j$ are connected by an edge with probability $p(n)$, independently. Given a bounded feature distribution $\Dcal$, let $\Gcal_\Dcal(n,c/n)$ denote the corresponding distribution on featured graphs.

We single out two cases of $p$ for {\erdosrenyi}:
\begin{itemize}
\item \emph{Dense}: $p$ is a constant.
\item \emph{Linear sparse}: $p$ is $\frac{c}{n}$ for some $c$.
\end{itemize}

\par

\paragraph{A real-valued logic with averaging operators}
 We now present the real-valued logics which we will analyze.
 
\begin{definition}[Averaging logic] 
    $\Aggo{\Mean, \LMean, \supl}$ is a term language which contains node variables $u, v, w, \ldots$ and terms defined inductively as follows.
    \begin{itemize}
        \item The \emph{basic terms}, or \emph{atomic terms}, are the node feature functions $\val_i(u)$ 
        for each node variable $u$ and feature function $\val_i$, constants $c$, the characteristic function of the edge relation $\mathrm E(u, v)$ and equality of nodes $u=v$.
        \item Given a term $\tau(\bar v, u)$ the \emph{global mean} for node variable $u$ is:
        \begin{equation*}
            \Mean_u \tau(\bar v, u)
        \end{equation*}
        And given a term $\tau(\bar w, v, u)$ the \emph{local mean} for node variables $u,v$ is:
        \begin{equation*}
            \LMean_{v E u} \tau(\bar w, v, u)
            \end{equation*}
        \item Given a term $\tau(\bar v, u)$ the \emph{global supremum} for node variable $u$ is:
        \begin{equation*}
            \sup_u \tau(\bar v, u)
        \end{equation*}
        \item Terms are closed under applying a  function symbol for each Lipschitz continuous $F \colon \R^m \to \R$.
    \end{itemize}
\end{definition}

We define the \emph{rank}
$\rank(\tau)$ 
of a term $\tau(\bar u)\in \Aggo{\Mean, \LMean,  \supl}$ as the maximum number of nested aggregators in $\tau$. 
Similarly, we define the \emph{supremum rank} $\srank(\tau)$ of $\tau$
as the maximum number of nested $\sup$ operators in $\tau$, and 
the \emph{mean rank} $\mrank(\tau)$ as the maximum number of nested $\Mean$ or $\LMean$ operators. 

\begin{definition}[Interpretation of terms]
    Let $\tau$ be a term with free variables $u_1, \ldots, u_k$. Let $\GG = (G,\bar u, \chi)$ be a multi-rooted featured graph with $\abs{\bar u} = k$. The \emph{interpretation} $\dsb{\tau}_{\GG}$ of a term $\tau$ in $\GG$ is defined recursively as follows:
    \begin{itemize}
        \item $\dsb{c}_{\GG} = c$ for any constant $c$.
        \item $\dsb{\val_i(u_{j})}_{\GG}  = \chi_i(u_j)$, the $i^{th}$ feature of the node $u_j$.
        \item $\dsb{\mathrm E(u_{i}, u_{j})}_{\GG}$ is $1$ when $(u_i, u_j) \in E(G)$ and $0$ otherwise, and similarly for equality.
        \item $\dsb{F(\tau_1, \ldots, \tau_m)}_{\GG}  = F(\dsb{\tau_1}_{\GG}, \ldots, \dsb{\tau_m}_{\GG})$ for any Lipschitz function $F$.
        \item Define $\dsb* {\Mean_v \tau}_{\GG}$ as:
        \begin{equation*}
            \frac 1 {\abs{V(G)}}
                \sum_{v \in V(G)} \dsb {\tau}_{\GG[v]}
        \end{equation*}
        \item Define $\dsb* {\LMean_{v E u_j} \tau}_{\GG}$ as:
        \begin{equation*}
            \frac 1 {\abs{\Ne(u_j)}}
                \sum_{v \in \Ne(u_j) } \dsb {\tau}_{\GG[v]}
        \end{equation*}
        if the denominator is nonzero, and zero other otherwise.
        \item $\dsb* {\sup_v \tau}_{\GG} = \max_{v \in V(G)} \dsb {\tau}_{\GG[v]}$.
    \end{itemize}
    Below we also use the notation $\dsb{\tau(\bar u)}_{(G, \chi)}$ for $\dsb{\tau}_{\GG}$.
\end{definition}

We use $\dsb{\tau}_{\Gcal_\Dcal(n, p)}$ to denote the random variable obtained by sampling $\GG$ from $\Gcal_\Dcal(n, p)$.



\paragraph{Examples of the term language}
Our term language is quite expressive.
\begin{itemize}
\item For any first-order logic graph formula $\phi(\bar u)$ (i.e. based on equality and the graph relation) there is a term $\tau_\phi(\bar u)$ that returns $1$ when $\phi$ holds and $0$ otherwise. When $\phi$ has no free variables, we refer to these as
\emph{first-order logic graph terms}.
We form $\tau_\phi$ inductively, applying global supremum to simulate existential quantifications, and using Lipschitz functions that extend the Boolean functions
$\wedge$, $\vee$, and $\neg$.
\item For any first-order logic sentence $\phi$, there is a term $\tau_{\% \phi}$  that returns the percentage of nodes
satisfying $\phi$. If we consider graphs with a feature function $\val_1(u)$ we can write a term that returns the average of $\val_1(u)$ on any graph satisfying $\phi$, and zero on any other graph.
\item If $\phi(u)$ is a first-order formula, then we can write a term $\tau(v)$ that returns
the percentage of $v$'s neighbors that satisfy $\phi$, returning zero if $v$ has no neighbors.
\end{itemize}

\paragraph{Prior convergence results for {\erdosrenyi}}
We will present results showing that terms in our language converge on random featured graph models that are based on {\erdosrenyi}. Our term language contains the characteristic functions of first-order logic sentences over ordinary graphs (without features), and our results will always extend prior logical convergence laws for first-order logic on such graphs.
Thus we summarize the relevant convergence results for first-order logic sentences over ordinary graphs.

In the dense graph case, \cite{faginzeroone,glebskii} showed that the probability of each first-order sentence goes to $0$ or $1$. For $\sup$-free terms, \cite{usneurips24} showed
that we have the stronger  convergence  in probability for the dense and linear sparse cases. For first-order logic graph terms,  convergence in probability only holds when you have a zero-one law. Our first main result will be for the dense case: we will show convergence in probability for our term language over random featured graphs based on dense {\erdosrenyi}: this is a common generalization of \cite{faginzeroone,glebskii} and \cite{usneurips24}.

For the ``root growth'' case ---  $n^{-\alpha}$ for $\alpha$  in $(0,1)$ ---
\cite{shelahspencersparse} showed that the probability does \emph{not} converge if $\alpha$ is rational, while for $\alpha$ irrational we have a zero-one law. 
Since our term language extends first-order logic, it follows that we \emph{cannot have} convergence even in distribution for
$n^{-\alpha}$ for $\alpha$ rational in $(0,1)$. In contrast, \cite{usneurips24} showed convergence in probability for $\sup$-free terms (only averaging).
We leave all of the root growth cases open here.

We now turn to the linear sparse and sublinear sparse ($O(\frac{1}{n^\beta})$ for $\beta >1$) cases. Here \cite{lynchlinearsparse} showed that the probability converges but does not have a zero-one law. 
Thus  our term language \emph{cannot have} convergence in probability
for the linear sparse or sublinear sparse case.
In the second main result in this paper,  we will show convergence in distribution for our term language in the linear sparse case We will be able to show an ``almost sure aggregate elimination'' result in this case, and then via an analysis of aggregate-free terms show convergence in distribution.
Although we do not deal explicitly with the sublinear case in this paper, we believe the same techniques and results apply there: see the discussion in Section \ref{sec:conc}.

\section{Convergence in Probability for Dense \texorpdfstring{\erdosrenyi}{Erdős–Rényi}} \label{sec:dense}

We begin with the simpler of our random featured graph models, i.e.\@ \erdosrenyi\ graphs with constant probability $p$.
We will show that each term in $\Aggo{\Mean, \LMean, \supl}$ converges in probability.

\begin{definition}
    Given a graph $G$ and a tuple of nodes $\bar u$, let $\GrTp(\bar u)$ be the set of quantifier-free formulas in the language of graphs (without node features) which hold of $\bar u$, its \emph{type}. For any $k$, let $\GrTpSp_k$ be the set of types in $k$ free variables. For $t(\bar u) \in \GrTpSp_k$, let $\Ext(t)$ be the set of one-variable extensions of $t$, and for $u_i \in \bar u$ let $\Ext_{u_i}(t)$ be the set of one-variable extensions of $t$ which have an edge to $u_i$.
\end{definition}

\begin{theorem}
    \label{thm:mean sup dense convergence}
    For every closed term $\tau$ in the language $\Aggo{\Mean, \LMean, \supl}$, the value $\dsb\tau_{\Gcal_\Dcal(n, c)}$ converges in probability.
\end{theorem}

The result above for closed terms will follow from an ``aggregate elimination theorem'' for terms that may have arbitrary free variables:

 \begin{theorem}[Aggregate elimination in the dense case] \label{thm:denseaggelim} Take a term $\tau$ with $k$ free variables and a graph quantifier-free type $t \in \GrTpSp_k$. Then there is a Lipschitz function $\lam_\tau^t \colon \FeatSp^k \to \R$ such that $\tau$ and $\lam_\tau^t$ ``agree in probability for inputs satisfying $t$''. That is, for every $\margin, \smallprob > 0$ and $n \in \N$ large enough, with probability at least $1 - \smallprob$ when sampling $\GG = (G, \chi)$ from $\Gcal_\Dcal(n, c)$, we have that for all tuples $\bar u$ of nodes:
    \begin{equation}
        \label{eq:ih; proof:mean sup dense convergence}
        \abs*{
            \doublesqbracket {\tau}_{\GG} 
            - 
            \lam_\tau^{\GrTp(\bar u)}(\chi(\bar u))
        } < \margin
        \tag{IH}
    \end{equation}
\end{theorem}

We will refer to the $\lam_\tau^t$ as the \emph{controllers} for $\tau$.
Clearly \cref{thm:mean sup dense convergence} follows directly from this, since
for a closed term $\tau$ the result requires  $\lam_\tau$ to be a constant function.

Before proving Theorem \ref{thm:denseaggelim}, we note the following lemma regarding the probability of  being close to the supremum of a function: see Appendix \ref{app:supbound} for the short proof.

\begin{lemma}
    \label{lem:probability of being close to sup}
    Let $X, Y$ be compact Euclidean domains, take $f \colon X \times Y \to \R$ Lipschitz continuous, and let $\mathcal C$ be a distribution with support $Y$. Then for every $\margin > 0$:
    \begin{equation*}
        \inf_{x \in X}
        \Pr_{y \sim \mathcal C}\left(
            f(x, y) \geq \sup_{y' \in Y} f(x, y') - \margin
        \right)
        > 0
    \end{equation*}
\end{lemma}

\begin{proof}[Proof of \cref{thm:denseaggelim}]
    We give the inductive definition of the controllers and the inductive proof that they satisfy the requirements of the theorem \eqref{eq:ih; proof:mean sup dense convergence} in parallel:

        
        \textbf{Constant case: $\tau \equiv c$.} Note that there is only one graph type $t$ on $0$ variables and that $\lam_\tau^t$ takes no arguments. Let $\lam_\tau^t \coloneqq c$. Then \eqref{eq:ih; proof:mean sup dense convergence} is immediate.

        
        \textbf{Value case: $\tau \equiv \val_i(u_{j})$.}
        Note that again there is only one graph type $t$ on $1$ variable. This time $\lam_\tau^t$ takes a single argument. Then define:
        \begin{equation*}
            \lam_\tau^t(\bar x) \coloneqq x_i
        \end{equation*}
        the $i^{th}$ element of the the tuple $\bar x$.

        
        \textbf{Edge relation case: $\tau \equiv \mathrm E(u, v)$.} Note that any $t \in \GrTpSp_2$ specifies whether there's an edge between the two nodes. So we let $\lam_\tau^t$ be either $1$ or $0$ depending on this.


        \textbf{Equality case: $\tau \equiv x = y$.} This is similar to the edge relation case.

        
        \textbf{Function application step: $\tau \equiv F(\pi_1, \ldots, \pi_m)$.} Given any type $t$, let $t \restriction \pi_i$ be the restriction of $t$ to the free variables of $\pi_i$. The controller $\lam_{\pi_i}^{t \restriction \pi_i}$ may take fewer arguments than $\lam_\tau^t$ will (as $\pi_i$'s free variables are a subset of $\tau$'s). However for convenience we abuse notation and allow $\lam_{\pi_i}^{t \restriction \pi_i}$ to take arguments for each free variable in $\tau$.
        
        Define:
        \begin{align*}
            \lam_\tau^t(\tuplevec x)
            \coloneqq 
            F(\lam_{\pi_1}^{t \restriction \pi_1}(\tuplevec x), \ldots, \lam_{\pi_m}^{t \restriction \pi_m}(\tuplevec x))
        \end{align*}
        As a Lipschitz function of Lipschitz functions this is Lipschitz.

        To prove \eqref{eq:ih; proof:mean sup dense convergence}, take $\margin, \smallprob > 0$. By applying \eqref{eq:ih; proof:mean sup dense convergence} for each $\pi_i$ and taking a union bound, for large enough $n$, with probability at least $1 - \smallprob$, for each $i$ and for all tuples $\bar u$ of nodes:
        \begin{equation*}
            \abs*{
                \doublesqbracket {\pi_i(\bar u)}_{(G, \chi)} - \lam_{\pi_i}^{\GrTp(\bar u)}(\chi(\bar u))
            } < \margin
        \end{equation*}
        Under this event:
        \begin{equation*}
            \abs*{
                \doublesqbracket {\tau(\bar u)}_{(G, \chi)} - \lam_\tau^{\GrTp(\bar u)}(\chi(\bar u))
            } < L_F \margin
        \end{equation*}
        where $L_F$ is the Lipschitz constant for $F$.


        \textbf{Global mean step: $\tau \equiv \Mean_v \pi$.} First take any $t(\bar u) \in \GrTpSp_k$ and $t'(\bar u, v) \in \Ext(t)$. As an extension type, $t'$ specifies which edges exist between $\bar u$ and $v$. Let $r(t')$ be the number of such edges. Define:
        \begin{equation*}
            \alpha(t, t') \coloneqq p^{r(t')} (1-p)^{k - r(t')}
        \end{equation*}
        Given any $\bar u$ which satisfies $t$, this is the expected proportion of nodes $v$ such that $\concat{\bar u}{v}$ satisfies $t'$.

        Now define:
        \begin{align*}
            \lam_\tau^t(\tuplevec x)
            \coloneqq \sum_{t' \in \Ext(t)} \alpha(t, t') \Ex_{\bar y \sim \Dcal} \left[\lam_\pi^{t'}(\tuplevec x, \bar y)\right]
        \end{align*}

        Take $\margin, \smallprob > 0$. By \eqref{eq:ih; proof:mean sup dense convergence} for $\pi$ there is $N_1$ such that for all $n \geq N_1$, with probability at least $1 - \smallprob$, for each $i$ and for all tuples $(\bar u, v)$ of nodes:
        \begin{equation}
            \label{eq:mean ih; proof:mean sup dense convergence}
            \abs*{
                \doublesqbracket {\pi(\bar u, v)}_{(G, \chi)} - \lam_{\pi}^{\GrTp(\bar u, v)}(\chi(\bar u, v))
            } < \margin
        \end{equation}

        Given any tuple $\bar u$, letting $t = \GrTp(\bar u)$ and $\dsb{t'(\bar u)} \coloneqq \{v \in V(G) \mid t'(\bar u, v)\}$ we can write:
        \begin{gather}
            \nonumber
            \doublesqbracket {\tau(\bar u)}_{(G, \chi)}
            =
            \frac 1 n
            \sum_v
                \doublesqbracket {\pi(\bar u, v)}_{(G, \chi)} \\
            \quad=
            \sum_{t' \in \Ext(t)} 
                \frac{\abs{\dsb{t'(\bar u)}}} n
                \left(
                    \frac 1 {\abs{\dsb{t'(\bar u)}}}
                    \sum_{v \in \dsb{t'(\bar u)}}
                        \doublesqbracket {\pi(\bar u, v)}_{(G, \chi)}
                \right)
            \label{eq:mean rewrite; proof:mean sup dense convergence}
        \end{gather}
        Now, each $\abs{\dsb{t'(\bar u)}}$ is a binomial random variable with parameter $\alpha(t, t')$.
        By Hoeffding's Inequality (Appendix~\ref{app:hoeffding}) and a union bound there is $N_2$ such that for all $n \geq N_2$, with probability at least $1-\smallprob$, for every $t \in \GrTpSp$ and $t' \in \Ext(t)$, and for every tuple $\bar u$ such that $t(\bar u)$ we have that:
        \begin{equation}
            \label{eq:mean proportion close; proof:mean sup dense convergence}
            \abs*{
                \frac {\abs{\dsb{t'(\bar u)}}} n
                -
                \alpha(t, t')
            }
            < \margin
        \end{equation}

        In this case, we have in particular that:
        \begin{equation*}
            \abs{\dsb{t'(\bar u)}}
            >
            (\alpha(t, t') - \margin) n
        \end{equation*}
        Since $p \in (0,1)$ we have that $\alpha(t, t') > 0$, so for small enough $\margin$ we have that $\alpha(t, t') - \margin > 0$.
        By Hoeffding's Inequality again and a union bound, there is $N_3 \geq N_2$ such that for all $n \geq N_3$, with probability at least $1 - \smallprob$, for all tuples $\bar u$, letting $t = \GrTp(\bar u)$, for all $t' \in \Ext(t)$ we have that:
        \begin{equation}
            \def\arraystretch{2}
            \left|
                \begin{array}{c}
                \displaystyle\frac 1 {\abs{\dsb{t'(\bar u)}}}
                \sum_{v \in \dsb{t'(\bar u)}}
                    \lam_{\pi}^{t'}(\chi(\bar u, v)) \\
                \displaystyle- \Ex_{\bar y \sim \Dcal}\left[\lam_{\pi}^{t'}(\chi(\bar u), \bar y)\right]
                \end{array}
            \right|
            <
            \margin
          \label{eq:sub-mean close; proof:mean sup dense convergence} 
        \end{equation}

        Putting it all together, considering the rewriting \eqref{eq:mean rewrite; proof:mean sup dense convergence} of $\doublesqbracket {\tau(\bar u)}_{(G, \chi)}$ and the definition of $\lam_\tau^t(\tuplevec x)$, and using \eqref{eq:mean ih; proof:mean sup dense convergence}, \eqref{eq:mean proportion close; proof:mean sup dense convergence} and \eqref{eq:sub-mean close; proof:mean sup dense convergence}, for $n \geq \max(N_1, N_2, N_3)$ with probability at least $1-3\smallprob$, for all tuples $\bar u$, letting $t = \GrTp(\bar u)$ we have that (noting that $\abs{\Ext(t)} = 2^k$):
        \begin{equation*}
            \abs*{
                \doublesqbracket {\tau(\bar u)}_{(G, \chi)}
                -
                \lam_{\tau}^{t}(\chi(\bar u))
            }
            <
            2^k\margin^2
        \end{equation*}
        Since we can control $2^k\margin^2$, this proves \eqref{eq:ih; proof:mean sup dense convergence} for $\tau$.


        \textbf{Local mean step: $\tau \equiv \LMean_{vEu_j} \pi$.} We proceed similarly to the global mean case. Given any $t(\bar u) \in \GrTpSp_k$ and $t'(\bar u, v) \in \Ext_{u_j}(t)$, let $r_{u_j}(t')$ be the number of edges to the new node, excluding the one from the node $u_j$. Define:
        \begin{equation*}
            \alpha_{u_j}(t, t') \coloneqq p^{r_{u_j}(t')} (1-p)^{k - 1 - r_{u_j}(t')}
        \end{equation*}
        We can then define the controller:
        \begin{align*}
            \lam_\tau^t(\tuplevec x)
            \coloneqq \sum_{t' \in \Ext_{u_j}(t)} \alpha_{u_j}(t, t') \Ex_{\bar y \sim \Dcal} \left[\lam_\pi^{t'}(\tuplevec x, \bar y)\right]
        \end{align*}

        The proof that \eqref{eq:ih; proof:mean sup dense convergence} holds proceeds as before. The only difference is that in order to apply our Hoeffding concentration argument, we need that for each tuple $\bar u$ the neighborhood size of $u_j$ is sufficiently large. For any $M \in \N$, by applying another concentration argument, there is $N_M \in \N$ such that for all $n \geq N_M$ we have that with probability at least $1 - \smallprob$, for all tuples $\bar u$ we have that $\abs{\Ne(u_j)} > M$. Choosing a sufficiently large $M$ and conditioning on this event allows us to proceed with our concentration arguments to prove \eqref{eq:ih; proof:mean sup dense convergence}.

        
        \textbf{Supremum step: $\tau = \sup_y \pi$.} Define $t \in \GrTp_k$:
        \begin{equation*}
            \lam_\tau^t(\tuplevec x) \coloneqq
            \max_{t' \in \Ext(t)}
                \;
                \sup_{\bar y \in \FeatSp}
                    \lam_\pi^{t'}(\tuplevec x, \bar y)
        \end{equation*}
        Note that the supremum is finite because $\FeatSp$ is bounded and $\lam_\pi^{t'}$ is Lipschitz.

        To see that $\lam_\tau^t$ is Lipschitz, it suffices to show that each $\sup_{\bar y \in \FeatSp} \lam_\pi^{t'}(\tuplevec x, \bar y)$ is Lipschitz, since the maximum of a finite number of Lipschitz functions is Lipschitz. For this, take $\tuplevec x, \tuplevec x' \in \FeatSp^k$ and fix $\gamma > 0$. There is $\bar y^* \in \FeatSp$ such that:
        \begin{equation*}
            \sup_{\bar y \in \FeatSp} \lam_\pi^{t'}(\tuplevec x, \bar y) - \gamma
            \leq 
            \lam_\pi^{t'}(\tuplevec x, \bar y^*)
        \end{equation*}
        Then:
        \begin{align*}
            &\sup_{\bar y \in \FeatSp} \lam_\pi^{t'}(\tuplevec x, \bar y)
            -
            \sup_{\bar y \in \FeatSp} \lam_\pi^{t'}(\tuplevec x', \bar y) \\
            &\qquad\leq
            \lam_\pi^{t'}(\tuplevec x, \bar y^*)
            -
            \sup_{\bar y \in \FeatSp} \lam_\pi^{t'}(\tuplevec x', \bar y)
            + \gamma \\
            &\qquad\leq
            \lam_\pi^{t'}(\tuplevec x, \bar y^*)
            -
            \lam_\pi^{t'}(\tuplevec x', \bar y^*)
            + \gamma \\
            &\qquad\leq
            L \norm*{(\tuplevec x, \bar y^*) - (\tuplevec x', \bar y^*)} + \gamma \\
            &\qquad=
            L \norm*{\tuplevec x - \tuplevec x'} + \gamma
        \end{align*}
        where $L$ is the Lipschitz constant of $\lam_\pi^{t'}$. Exchanging the roles of $\tuplevec x$ and $\tuplevec x'$ gives:
        \begin{equation*}
            \abs*{
                \sup_{\bar y \in \FeatSp} \lam_\pi^{t'}(\tuplevec x, \bar y)
                -
                \sup_{\bar y \in \FeatSp} \lam_\pi^{t'}(\tuplevec x', \bar y)
            }
            \leq
            L \norm*{\tuplevec x - \tuplevec x'} + \gamma
        \end{equation*}
        Finally using that $\gamma > 0$ was arbitrary, we have that $\sup_{\bar y \in \FeatSp} \lam_\pi^{t'}(\tuplevec x, \bar y)$ is Lipschitz.

        Now take $\margin, \smallprob > 0$. By \eqref{eq:ih; proof:mean sup dense convergence} for $\pi$ there is $N_1$ such that for all $n \geq N_1$, with probability at least $1 - \smallprob$, for all tuples $(\bar u, v)$ we have:
        \begin{equation*}
            \label{eq:sup ih; proof:mean sup dense convergence}
            \abs*{
                \doublesqbracket {\pi(\bar u, v)}_{(G, \chi)} - \lam_{\pi}^{\GrTp(\bar u, v)}(\chi(\bar u, v))
            } < \margin
        \end{equation*}

        Given any tuple $\bar u$, letting $t = \GrTp(\bar u)$ we can write:
        \begin{align*}
            \doublesqbracket {\tau(\bar u)}_{(G, \chi)}
            &=
            \max_v
                \doublesqbracket {\pi(\bar u, v)}_{(G, \chi)} \\
            \label{eq:sup rewrite; proof:mean sup dense convergence}
            &=
            \max_{t' \in \Ext(t)} 
                \;
                \max_{v \in \dsb{t'(\bar u)}}
                    \doublesqbracket {\pi(\bar u, v)}_{(G, \chi)}
        \end{align*}

        Using a Hoeffding concentration argument as above, for any $M > 0$ there is $N_M$ such that for all $n \geq N_M$, with probability at least $1 - \smallprob$, for all tuples $\bar u$, letting $t = \GrTp(\bar u)$, for all $t' \in \Ext(t)$ we have that:
        \begin{equation*}
            \abs{\dsb{t'(\bar u)}} > M
        \end{equation*}

        Take any $t \in \GrTpSp_k$ and $t' \in \Ext(t)$. We need to show that for all $\tuplevec x \in \FeatSp^k$:
        \begin{equation*}
            \max_{v \in \dsb{t'(\bar u)}} \lam_\pi^{t'}(\tuplevec x, \chi(v))
        \end{equation*}
        is close to:
        \begin{equation*}
            \sup_{\bar y \in \FeatSp} \lam_\pi^{t'}(\tuplevec x, \bar y)
        \end{equation*}
        Let $q_{t'}$ be the supremum for $\tuplevec x \in \FeatSp^k$ of:
        \begin{align*}
            \Pr_{\bar z \sim \Dcal} \left(
                \lam_\pi^{t'}(\tuplevec x, \bar z)
                \leq
                \sup_{\bar y \in \FeatSp} \lam_\pi^{t'}(\tuplevec x, \bar y)
                - 
                \margin
            \right)
        \end{align*}
        We have that $q_{t'} < 1$ by Lemma \ref{lem:probability of being close to sup}.

        Now, we have fixed a tuple $\bar u$. Formally, we should imagine generating a featured graph structure on a fixed sequence of $n$ nodes, so that `fixing $\bar u$' means `fixing a tuple of node indices'. We wish to consider the nodes in $\dsb{t'(\bar u)}$. Formally, this is a random subset of the node indices, where, because edges are sampled independently in the \erdosrenyi\ distribution, each node index is sampled independently. Each node has feature distribution sampled from $\Dcal$ independently. Therefore, if we condition on $\dsb{t'(\bar u)}$ having size $J$, taking a union bound, the probability that:
        \begin{equation*}
            \abs*{
                \max_{v \in \dsb{t'(\bar u)}} \lam_\pi^{t'}(\chi(\bar u), \chi(v))
                -
                \sup_{\bar y \in \FeatSp} \lam_\pi^{t'}(\chi(\bar u), \bar y)
            }
            >
            \margin
        \end{equation*}
        is at most $\left(q_{t'}\right)^J$.
        
        Since $q_{t'} < 1$ can choose $M$ large enough so that for all $t'$ we have that:
        \begin{equation*}
            \left(q_{t'}\right)^M < \smallprob
        \end{equation*}

        Then, for every $n \geq N_M$, it holds with probability at least $1 - (2^k + 1)\smallprob$ that
        for all tuples $\bar u$ we have:
        \begin{equation*}
            \abs*{
                \doublesqbracket {\tau(\bar u)}_{(G, \chi)}
                -
                \lam_{\tau}^{t}(\chi(\bar u))
            }
            <
            2\margin
        \end{equation*}
        This proves \eqref{eq:ih; proof:mean sup dense convergence} for $\tau$. \qedhere
\end{proof}

\section{Convergence in Distribution for Linear Sparse \texorpdfstring{\erdosrenyi}{Erdős–Rényi}} \label{sec:linearsparse}

We now consider the case of random featured graphs based on \erdosrenyi\ where the edge probability $p(n)=\frac{c}{n}$ for $c>0$ a constant. We \emph{fix $c$ for this section}, and restrict to $n$ large enough that $\frac{c}{n} \leq 1$, referring to the corresponding random graph model as \emph{linear sparse}. We can assume that our MRFGs all take values in $\FeatSp$, and we do this throughout the section.

Recall from Section \ref{sec:prelims} that for first-order logic over graphs, we have
convergence of probabilities in this model: for each first-order sentence $\psi$ its probability $P_n(\psi)$ converges.

We do not have a zero-one law for first-order logic, so consider first-order logic sentence $\psi$ such that $P_n(\psi)$ converges to $r$ with $0<r<1$. Since our term language includes the characteristic
function $\chi_\psi$ of $\psi$, we have a term that is $0$ for $r$ percentage of the graphs, asymptotically, and $1$ for $1-r$ of the graphs. Thus \emph{we cannot hope to prove convergence in probability for each term in our language}.

The main result of this section is:

\begin{theorem} \label{thm:mainthmlinearsparse} 
    For every closed term $\tau$ in the language $\Aggo{\Mean, \LMean, \supl}$, the value $\dsb\tau_{\Gcal_\Dcal(n, c/n)}$ converges in distribution.
\end{theorem}


This theorem generalizes the convergence law for first-order logic on $\Gcal(n,c/n)$ shown in \cite{lynchlinearsparse}. To gain a better understanding of that work and ours it is useful to informally describe the ``local'' landscape of $\Gcal(n,c/n)$ \cite{shelah1994can,van_der_Hofstad_2024_Vol2}. Fix an integer $r>0$. Then a.a.s.\@ (1)
all $r$-neighborhoods $\ball_r(v)$ are either trees or unicycles, (2) there are ``few'' unicyclic $r$-neighborhoods, and they are far apart, and (3) the neighborhood $\ball_r(v)$ obtained by sampling a vertex $v$ uniformly at random is similar to a branching process with Poisson offspring distribution (see Subsection~\ref{subsec:linear_prelims} for a definition). Globally, $\Gcal(n,c/n)$ has a much more complex structure. However, first-order logic of a fixed quantifier rank $k\geq 0$ is, in some sense, oblivious to phenomena that cannot be detected in neighborhoods of radius $r= O(3^k)$ \cite{gaifmanlocality}. \par

The key idea in \cite{lynchlinearsparse} is to define, via a game, a notion of similarity on graphs that partitions graphs into finitely many classes, and to show:
\begin{enumerate} 
    \item (\emph{A.a.s.\@ simplification}) Every formula is, a.a.s., a union of similarity classes.
    \item These similarity classes can be characterized purely graph-theoretically --- they are determined
    by the union of $r$-neighborhoods of all cycles of size up to $r$,
    for suitable $r$, where this union
    of neighborhoods is called the ``$r$-core'' of the graph.
\end{enumerate}
Then using the graph-theoretic characterization based on cores, one can infer:
\begin{enumerate}[resume*]
    \item On each similarity class we have convergence in the probability for formulas.
\end{enumerate}
Combining 1) and 3) gives the final convergence result.

Here we apply a similar approach. We develop a notion of similarity on featured graphs, again via a kind of game,  which we show is an appropriate analog of standard pebble games for a term language with supremum, Lipschitz functions, and local averaging. Unlike \cite{lynchlinearsparse},
our notion of similarity is not an equivalence relation. The games will have ``accuracy parameters'' which measure, for example, how big of a difference between features we consider admissible. 
We show:
\begin{enumerate}
    \item (\emph{A.a.s.\@ simplification}) A.a.s.\@ every term in our language reduces to a ``global-mean--free'' expression: one using only local averaging. In analogy to what we did in the dense case, we call these expressions \emph{controllers}. From this it will follow that the value of terms in the language are determined, up to a granularity measure within the feature space,  by the similarity class of the featured graphs. 
    \item Featured graph similarity can be related to an adaptation of the notion of $r$-core to featured graphs. 
\end{enumerate}
We use the connection of similarity with the $r$-core to show: 
\begin{enumerate}[resume*]
    \item The global-mean-free controllers appearing in the a.a.s.\@ simplification converge in distribution. 
\end{enumerate}
Combining the first  and third items provides our final convergence result.

Let us give an overview of this section and a roadmap for the proof of Theorem \ref{thm:mainthmlinearsparse}.

\begin{itemize}
\item In Subsection \ref{subsec:linear_prelims} we give additional preliminaries for the rest of the section.
\item In Subsection \ref{subsec:gamesdiscret} we introduce some pebble games that extend so-called Ehrenfeucht-Fra\"isse games \cite{fmtbook} that characterize the expressive power of first-order logic. 
\item In Subsection \ref{subsec:linear_controllers} we define, for each term $\tau$, a \emph{controller} expression $\lambda_\tau$ that includes no global averaging operators. The main results of this section are that controllers only depend on the cores of MRFGs (Lemma \ref{lem:controller_core}), and that controllers take similar values on pairs of MRFGs that cannot be distinguished through our pebble games (Theorem \ref{thm:controller_vs_controller_approx}).
\item In Subsection \ref{subsec:modeltheoretic} we will present some \emph{axioms}, representing properties that hold in typical featured graphs. The main result in that subsection, Theorem \ref{thm:aggelimsparse}, shows that on MRFGs satisfying these axioms, each term $\tau$ is close to its controller $\lambda_\tau$. 
\item In Subsection \ref{subsec:combinatorial}, the ``combinatorial part'', we show that a.a.s. $\Gcal_\Dcal(n,c/n)$ satisfies the axioms laid out in the previous subsection. In consequence, the value of a term $\tau$ converges in probability to the value of its controller $\lambda_\tau$ (Corollary \ref{cor:convergence_in_probability}).
\item Finally, in \ref{subsec:linear_main_proof} we put everything together and prove our main result, Theorem \ref{thm:mainthmlinearsparse}.
\end{itemize}

\subsection{Auxiliary Definitions}
\label{subsec:linear_prelims}

\paragraph{Slopes of functions and terms}
The slope $\slope_F$ of a (globally) Lipschitz function $F: X \rightarrow \R$
with $X\subseteq \R^m$ is the infimum value $d$ satisfying 
that:
\[
\abs*{F(\bar x) - F(\bar y)} \leq d \norm*{\bar x - \bar y}_\infty
\]
for all $\bar x , \bar y \in X$. 

We extend this to give a definition of the \emph{slope} $\slope_\tau$ of a term $\tau(\bar u)\in \Aggo{\Mean, \LMean, \supl}$ recursively as follows:
\begin{itemize}
    \item $\slope_\tau=1$ if $\tau(\bar u)\equiv \val(u_j)$ for some $j$,
    \item $\slope_\tau=1$ if $\tau(\bar u) \equiv \mathrm E(u_i, u_j)$ for some $i,j$,
    \item $\slope_\tau= \max_{i\in [m]} \{  
    \slope_F \cdot \slope_{\pi_i} \}$ if $\tau(\bar u) \equiv F(\pi_1, \dots, \pi_m)$,
    \item $\slope_\tau = \slope_{\pi}$ if $\tau(\bar u)\equiv \sup_v \pi(\bar u, v)$ and,
    \item $\slope_\tau = \slope_{\pi} + 1$ if  $\Mean_v \pi(\bar u, v), \mbox{ or }  \LMean_{v E u} \pi(\bar u, v)$.
\end{itemize}

\paragraph{Cores and disjoint unions}
Given an integer $r\geq 0$,
we write
$\coreof{\GG}{r}$ 
for the MRFG $\HH$ obtained by restricting $\GG$
to the vertices $v$ 
that are at distance at most $r$ to some root of $\GG$
or some cycle of size at most $2r+1$. We also call $\coreof{\GG}{r}$   the $r$-\emph{core} of $\GG$, as in \cite{lynchlinearsparse}. The intuition is that the $r$-core $\coreof{\GG}{r}$ contains all the ``interesting'' $r$-neighborhoods of $\GG$. The maximum size of a cycle with no cords that can fit inside a $r$-neighborhood is $2r+1$, so for all vertices $v$ outside $\coreof{\GG}{r}$, the neighborhood
$\ball_r(v)$ is a tree containing no root from $\GG$. 
\par
Given two MRFGs $\GG=(G, \bar u, \chi_G), \HH=(H, \bar v, \chi_H)$, we define their disjoint union, denoted $\GG\sqcup \HH$ as: $(G\sqcup H, \concat{\bar u}{\bar v}, \chi_{G\sqcup H})$ where $\chi_{G\sqcup H}(w)$ equals $\chi_{G}(w)$ if $w\in V(G)$, and equals $\chi_H(w)$ otherwise. 
Observe that $\GG$ may have $j$ roots while $\HH$ has $k \neq j$ roots, and $\GG \sqcup \HH$ will then have $j+k$ roots.  

\paragraph{Branching processes}
We define the \emph{branching  process} $\BP$
as the random rooted tree $(T,v)$ generated by letting the number of children of each vertex follow a Poison distribution $\Po_c$ with parameter $c$, independently for each vertex. Given an integer $r\geq 0$, we define $\coreof{\BP}{r}$ as the random rooted tree obtained by considering the first $r$ generations of $\BP$.
The \emph{featured branching process} $\FBP$ is a random rooted featured tree $(T,v,\chi)$, where $(T,v)$
follows the distribution $\BP$, and $\chi(u)$ follows the distribution $\Dcal$ independently for each
$u\in V(T)$. We define $\coreof{\FBP}{r}$ analogously to $\coreof{\BP}{r}$. Observe that $\coreof{\FBP}{r}$ is precisely the $r$-neighborhood of its root. In other words, $\coreof{\FBP}{r}$ is the $r$-core of $\FBP$, so the notation is consistent. 
\par

\paragraph{Random cores}

Given an integer $r\geq 3$, we define the ``random $r$-core'' denoted 
$\mathrm{Core}_{r}$, as the random graph obtained by generating $\Po_{c^i/2i}$ cycles of length $i$ independently for each $3\leq i\leq r$, and attaching a copy of $\BP\vert_r$ to each vertex lying on a cycle independently. 

In the last part of the proof of our main result (see Section \ref{subsec:linear_main_proof}) we will make use of the fact that the $r$-cores of sparse random graphs converge in distribution to $\mathrm{Core}_r$:

   \begin{fact}[Core convergence of sparse random graphs; Lemma 2.6 in \cite{larrauri2023thesis}]
\label{fact:core_tight_distribution}
    Let $r\geq 0$. 
    For each $n\geq 1$, let $H_n$ denote the $r$-core of the random graph $\Gcal(n,c/n)$. Then for each graph $G$, the limit $\lim_{n\to \infty} \Pr(H_n\simeq G)$ exists and is equal to $\Pr(\mathrm{Core}_r \simeq G)$.
\end{fact}
Similarly, we define $\mathrm{Core}_{r,\Dcal}$
as the random featured graph whose underlying graph is $\mathrm{Core}_{r}$,
and where the features of each vertex have distribution $\Dcal$ independently.

\paragraph{Couplings}
A coupling of two random variables $X$, $Y$ is a vector-valued random variable $\Pi= (\Pi_X, \Pi_Y)$ satisfying that $\Pi_X$ is distributed like $X$ and $\Pi_Y$ is distributed like $Y$. We also define the coupling of $X$ and $Y$ when $X$ or $Y$ is a set instead of a random variable. In this case we define the coupling for the uniform random variable over the set.

\subsection{Games} 
\label{subsec:gamesdiscret}

We introduce reflexive relations 
$\sim_{k,\epsilon,\eta}$ over the space of MRFGs
for each $\epsilon, \eta>0$, and each integer $k$. These ``similarity relations'' can be represented via games that capture closeness under the global-mean--free fragment of $\Aggo{\Mean, \LMean, \supl}$, the analog  of standard pebble games for first-order logic \cite{gradel2007fmt}. The relation $\sim_{k,\epsilon,\eta}$ is inductively defined
for each integer $k\geq 0$ as follows.
\begin{itemize}
    \item $(G,\bar v,\chi) \sim_{0,\epsilon,\eta} (H,\bar u, \nu)$ whenever $\bar v\mapsto \bar u$ is
    a partial isomorphism between $G$ and $H$, and $\lVert \chi(v_i) - \nu(u_i) \rVert_\infty \leq \epsilon$ for all $1\leq i \leq \abs{\bar v}$.
    \item If $k\geq 1$, then $(G,\bar v,\chi) \sim_{k,\epsilon,\eta} (H, \bar u, \nu)$ whenever
    the following two properties hold:
    \begin{enumerate}[label=(\arabic*)]
        \item \textit{Back-and-forth property.} For all $p\in V(G)$ there is $q\in V(H)$ such that 
        $(G,\bar v p,\chi)$ $\sim_{k-1, \epsilon,\eta} (H, \bar u q, \nu)$. Similarly, for all $q\in V(H)$ there is $p\in V(G)$ such that 
        $(G,\bar v p,\chi)$ $\sim_{k-1,\epsilon,\eta} (H, \bar u q, \nu)$.
        \item \textit{Neighborhood-coupling property.} For all $1\leq i\leq |\bar v|$ the $i^{th}$ root $v_i\in V(G)$ is isolated if and only if $u_i\in V(H)$ is isolated as well. Further,  if neither $v_i$ nor $u_i$ are isolated there is a coupling   $\Pi=(\Pi_G, \Pi_H)$ of $\Ne(v_i)$ and $\Ne(u_i)$ satisfying that:
        \begin{equation}
        \label{eq:coupling_game}
                \Pr_{(v,u)\sim \Pi}( \GG[v] \sim_{k-1,\epsilon,\eta} \HH[u])  \geq 1 - \eta.
        \end{equation}
    \end{enumerate}
\end{itemize}

A way to understand this relation is through a two-player game, played on the MRFGs $\GG$ and $\HH$. The players of this game are called Spoiler
and Duplicator. The goal of Spoiler is to show that $\GG$ and $\HH$ are ``very different'' from the perspective of $\Aggo{\LMean, \supl}$ (the $\Mean$-free fragment of $\Aggo{\Mean, \LMean, \supl}$), while Duplicator wants to argue that the MRFGs are similar. Let $\GG_0=\GG, \HH_0=\HH$. The game proceeds in $k$ rounds. At the end of the $i^{th}$ round vertices $v\in V(G)$ and $u\in V(H)$ are selected and added as roots, defining 
$\GG_i=\GG_{i-1}[v]$, $\HH_i=\HH_{i-1}[u]$. Spoiler wins if at any point
the map that matches the roots of $\GG_i$ to the roots of $\HH_i$ is not a partial isomorphism between $G$ and $H$, or if there is some root $v$ in $\GG_i$ whose features are very different from those of the corresponding root $u$ in $\HH_i$. More precisely, this occurs when $\lVert \chi(v) - \nu(u) \rVert_\infty > \epsilon$. Duplicator wins if by the end of the $k^{th}$ round Spoiler has not won. 

At the beginning of the $i^{th}$ round Spoiler makes one of two different kind of moves. In the first, he picks a vertex from either $\GG_i$ or $\HH_i$ and then Duplicator responds by picking a vertex in the other MRFG. This simulates the \emph{back and forth} property in $\sim_{k,\epsilon,\eta}$, and captures the behavior of the $\supl$ aggregator. In the second type of move, Spoiler picks corresponding roots $v\in V(G)$, $u\in V(H)$ and challenges Duplicator to prove that $\Ne(v)$ and $\Ne(u)$ are similar. Duplicator then replies by giving a coupling $\Pi$ of $\Ne(v)$ and $\Ne(u)$ and choosing a high-probability set
$S\subseteq \Ne(v) \times \Ne(u)$, which means, precisely, that
$\Pr(\Pi\in S)\geq 1 - \eta$. Then Spoiler chooses a pair $(v^\prime, u^\prime)\in S$ and the game continues. This simulates the \emph{neighborhood coupling} property in $\sim_{k,\epsilon,\eta}$, and captures the behavior of the $\LMean$ aggregator. Then, $\GG \sim_{k,\epsilon,\eta} \HH$ holds precisely when Duplicator wins this game. \par

A fact that we use repeatedly is that $\sim_{k,\epsilon,\eta}$ is preserved under disjoint unions. That is, $\GG_1\sim_{k,\epsilon,\eta} \HH_1$ and $\GG_2 \sim_{k,\epsilon,\eta} \HH_2$ imply that 
$\GG_1 \sqcup \GG_2 \sim_{k,\epsilon,\eta} \HH_1 \sqcup \HH_2$.
This can be shown through a ``strategy composition'' argument: Duplicator can win the game on the disjoint unions by playing according to winning strategies on each of the  disjoint parts.

\subsection{Controllers} \label{subsec:linear_controllers}

Given a term $\tau(\bar u)\in \Aggo{\Mean, \LMean, \supl}$, we define the \emph{controller function} $\lambda_\tau$ over MRFGs with $\abs{\bar u}$ roots. Let $\GG=(G,\bar u, \chi)$ be a MFRG. Then the
value $\lambda_\tau(\GG)$ represents the value of $\tau$ on the disjoint union of $\GG$ and an infinite featured forest that looks locally like $\FBP$. Formally, $\lambda_\tau(\GG)$
is defined inductively as follows:
\begin{itemize}
    \item When $\tau(\bar u)\equiv
    \val_i(u_j)$ let $\lambda_\tau(\GG)=\chi_i(u_j)$,
    \item When
    $\tau(\bar u)\equiv
    \mathrm E(u_i, u_j)$ let $\lambda_\tau(\GG)=1$ when $(u_i,u_j)\in E(G)$ and $0$ otherwise,
    \item    When $\tau(\bar u)\equiv
    F(\pi_1,\dots, \pi_m)$ define the controller as $\lambda_{\tau}(\GG)= F(\lambda_{\pi_1}(\GG),\dots, \lambda_{\pi_m}(\GG))$,

    \item When   $\tau(\bar u)=\Mean_v \pi(\bar u, v)$ let $\lambda_\tau(\GG)$ be:
    \[
    \Ex_{\TT\sim \FBP_{\Dcal}}
    \biggl[ 
    \lambda_{\pi}
    (\GG\sqcup \TT)
    \biggr],    
    \]
  
    \item When  $\tau(\bar u)\equiv \sup_v \pi(\bar u, v)$ let $\lambda_\tau(\GG)$ be:
    \begin{equation*}
        \max \left\{
            \sup_{v\in V(G)}
            \lambda_{\pi}(\GG[v]),  
            \sup_{\substack{\TT \text{ rooted}\\ \text{featured tree}}} 
            \lambda_{\pi}(\GG\sqcup \TT)
        \right\}
    \end{equation*}

    \item When $\tau(\bar u) \equiv \LMean_{v E u_i} \pi(\bar u, v)$ let
    $\lambda_{\tau}(\GG)$ be $0$
if $\GG$'s $i^{th}$ root is an isolated vertex, and otherwise:
    \[
    \frac{1}{|\Ne(u_i)|} \sum_{v\in \Ne(u_i)} \lambda_{\pi}(\GG[v]) 
    \]
\end{itemize}
Observe that, unlike the dense case, this time the controller functions contain expressions that are not part of the term language, both due to the $\Mean$ and $\supl$ constructions.
But since the controllers do not contain the $\Mean$ operator, they are preserved by the games:

\begin{theorem}[Preservation of controllers by games]
\label{thm:controller_vs_controller_approx}
    Let $\epsilon, C>0$.
    Let $\tau(\bar u)\in \Aggo{\Mean, \LMean, \supl}$
    be a term satisfying that 
    $| \lambda_{\tau^\prime} | \leq C$ for all sub-terms $\tau^\prime$ of $\tau$, and let $k \geq \srank(\tau) + \lmrank(\tau)$ be an integer. 
    Consider two MFRGs $\GG$ and $\HH$ with $|\bar u|$ roots. Suppose that  
    $\GG \sim_{k,\epsilon,\eta} \HH$, for $\eta= \frac{\epsilon}{4C}$.
    Then:
    \begin{equation}
    \label{eq:EF_controller}
            \Big\lvert \lambda_\tau(\GG) - \lambda_\tau(\HH) \Big\rvert \leq 
    \epsilon \cdot \slope_\tau. 
    \end{equation}
\end{theorem}

\begin{proof}   
    Fix $\epsilon, C>0, \eta= \frac{\epsilon}{4C}$. We show the result by induction on $\tau$'s structure. The statement is clearly true when $\tau$ is an atomic term. We deal with each induction step as follows: \par
     \textbf{Function application step: $\tau \equiv F(\pi_1, \dots, \pi_m)$.} Observe that
         $\srank(\pi_i) \leq \srank(\tau)$, and
         $\lmrank(\pi_i) \leq \lmrank(\tau)$ for all $i\in [m]$, so:
         \[
        \Big\lvert \lambda_{\pi_i}(\GG) - \lambda_{\pi_i}(\HH) \Big\rvert \leq 
    \epsilon \cdot \slope_{\pi_i}. 
         \]
         Hence:
         \begin{align*}
            \Big\lvert \lambda_\tau(\GG) - \lambda_\tau(\HH) \Big\rvert 
            &\leq 
            \max_{i\in [m]} \slope_F \cdot   \Big\lvert \lambda_{\pi_i}(\GG) - \lambda_{\pi_i}(\HH) \Big\rvert \\
            &\leq
            \max_{i\in [m]} \epsilon \cdot \slope_F \cdot \slope_{\pi_i} \\
            &\leq
            \epsilon \cdot\slope_{\tau}. 
        \end{align*}
    \par
    \textbf{Global mean step: $\tau \equiv \Mean_v \pi$.} By the induction hypothesis, 
    for any finite rooted featured tree $\TT$ we have that:
    \[
    \Big\lvert \lambda_{\pi}(\GG \sqcup \TT) - \lambda_{\pi}(\HH \sqcup \TT) \Big\rvert \leq 
    \epsilon \cdot \slope_{\pi}. 
    \]
    To see this, observe that $\srank(\pi) + \lmrank(\pi) \leq k$ and clearly
    $\GG\sqcup \TT \sim_{k,\epsilon,\eta}
    \HH\sqcup \TT $. Now 
     $\lvert \lambda_{\tau}(\GG) - \lambda_{\tau}(\HH) \rvert \leq 
    \epsilon \cdot \slope_{\tau}$ follows from the definition of $\lambda_\tau$ together with the fact that $\slope_\tau= \slope_{\pi} + 1$. \par
    \textbf{Supremum step: $\tau \equiv \sup_v \pi$.} In order to prove that the statement holds for $\tau$ it is enough to show that
    $\lambda_\tau(\GG) \leq \lambda_\tau(\HH) + \epsilon \cdot \slope_{\tau}$ 
    and 
     $\lambda_\tau(\HH) \leq \lambda_\tau(\GG) + \epsilon \cdot \slope_{\tau}$. We prove the first inequality, the second can be shown analogously. 
    There are two sub-cases corresponding to the definition of $\lambda_\tau$.
    \textbf{Sub-case 1.} Suppose there is some $v_G\in V(G)$ for which
    $\lambda_{\tau}(\GG)=
    \lambda_{\pi}(\GG[v_G])$.
    Let $v_H\in V(H)$ be a vertex
    satisfying $\GG[v_G] \sim_{k-1,\epsilon,\eta} 
    \HH[v_H]$. By hypothesis we have that:
        \[
    \Big\lvert \lambda_{\pi}(\GG[v_G]) - \lambda_{\pi}(\HH[v_H]) \Big\rvert \leq 
    \epsilon \cdot \slope_{\tau}, 
    \]
    using that $\slope_{\tau}=\slope_{\pi}$. Now
    $\lambda_\tau(\GG) \leq \lambda_\tau(\HH) + \epsilon \cdot \slope_{\tau}$ follows from the fact that
    $\lambda_{\pi}(\HH[v_H])\leq
    \lambda_{\tau}(\HH)$. 
    \textbf{Sub-case 2.} Suppose there is some finite featured rooted tree $\TT$ for which
    $\lambda_\tau(\GG)= \lambda_{\pi}(\GG \sqcup \TT)$. It holds that
    $\GG\sqcup \TT \sim_{k,\epsilon,\eta}
    \HH\sqcup \TT$. Hence, by assumption:
        \[
    \Big\lvert \lambda_{\pi}(\GG \sqcup \TT) - \lambda_{\pi}(\HH \sqcup \TT) \Big\rvert \leq 
    \epsilon \cdot \slope_{\tau}. 
    \]
    Again, now
    $\lambda_\tau(\GG) \leq \lambda_\tau(\HH) + \epsilon \cdot \slope_{\tau}$ follows from the fact that
    $\lambda_{\pi}(\HH\sqcup \TT)\leq
    \lambda_{\tau}(\HH)$. \par
    \textbf{Local mean step: $\tau \equiv \LMean_{v E u_i} \pi$.} Let $v_G, v_H$ denote the $i^{th}$ roots of $\GG$ and $\HH$ respectively. There are two sub-cases. \textbf{Sub-case 1.} Suppose that both $v_G$ and $v_H$ are isolated vertices. Then 
    $\lambda_\tau(\GG)=\lambda_\tau(\HH) = 0$, and the statement holds. \textbf{Sub-case~2.} Now suppose that both $v_G$ and $v_H$ are non-isolated. Let $\Pi$ be a coupling of $\Ne(v_G)$ and $\Ne(v_H)$ satisfying \eqref{eq:coupling_game} for $\eta= \frac{\epsilon}{4C}$. Then:
    \begin{align*}
        &\lvert \lambda_\tau(\GG) - \lambda_\tau(\HH) \rvert  \\ 
        &\quad= \abs*{
            \Ex_{(u_G,u_H)\sim \Pi}
            \left[  ~
            \lambda_{\pi}(\GG[u_G]) - \lambda_{\pi}(\HH[u_H]) ~
            \right]
        } \\
        &\quad\leq
        \Ex_{(u_G,u_H)\sim \Pi}\left[  
        ~ \lvert
        \lambda_{\pi}(\GG[u_G]) - \lambda_{\pi}(\HH[u_H])
        \rvert ~
        \right] \\    
        &\quad\leq
        \left(1 - \frac{\epsilon}{4C}\right) \slope_{\pi}\epsilon +
        \frac{\epsilon}{4C} 2C \\
         &\quad\leq (\slope_{\pi} + 1) \epsilon = \slope_{\tau} \epsilon.
    \end{align*}
    The first equality uses linearity of expectation, together with the fact that the marginal distributions of $u_G$ and $u_H$ are uniform over $\Ne(v_G)$ and $\Ne(v_H)$ respectively. 
    The second inequality uses the fact that
    $\GG[u_G] \sim_{k-1,\epsilon,\eta}\HH[u_H]$ with probability at least $1- \epsilon/4C$,
    and  $|\lambda_{\pi}|$ is bounded by $C$.\qedhere
\end{proof}

We note two additional properties of controllers, which apply also to terms that do not contain global mean.
One is that controller images are bounded: see Appendix \ref{app:controllersboundedimage} for the short inductive proof.
\begin{proposition}[Controllers have bounded image]
\label{prop:diam_controller}
    Let $\tau(\bar u) \in \Aggo{\Mean, \LMean, \supl}$ be a term. Then $\abs{\lambda_\tau}$ is bounded.
\end{proposition}

The second key property is that a controller value only depends on its $r$-core for suitable $r$ (see Appendix \ref{app:coredeterminacy}):

\begin{lemma}[Core determinacy of controllers]
\label{lem:controller_core}
Let $\tau(\bar u)\in \Aggo{\Mean, \LMean, \supl}$ be a term, and let $k=\srank(\tau) + \lmrank(\tau)$. Then, for each MRFG $\GG$ we have that:
\[
\lambda_\tau(\GG)=
\lambda_\tau(\GG\vert_{r_k}).
\]
\end{lemma}

\subsection{\texorpdfstring{``Model-Theoretic''}{"Model-Theoretic"} Part} 
\label{subsec:modeltheoretic}
We are now ready to give axioms that should hold in almost every graph. 
These will be sufficient to guarantee that a term $\tau$ simplifies to its controller $\lambda_\tau$. 
For the following definitions we consider parameters $k, \epsilon, \eta, r$, where $\epsilon,\eta> 0$, and $k,r\geq 0$ are integers. 
Recall the intuition that $\lambda_\tau(\GG)$ represents the result of evaluating $\tau$ on the disjoint union of $\GG$ and an infinite forest that locally looks like $\FBP$. 
The axioms reflect that $\GG$ itself is similar, from the perspective of $\sim_{k,\epsilon,\eta}$, to this disjoint union. The parameter $k$ will be $\rank(\tau)$ in applications, and $r$ will represent a bound on the radius of the neighborhoods that influence $\tau$, usually $r=\frac{3^k-1}{2}$. The parameters $\epsilon, \eta$ are error parameters that will control the convergence of $\lambda_\tau$ to $\tau$. \par

The following richness axiom is a variation of the richness property defined in \cite{lynchlinearsparse}, and states that for any rooted featured tree $\TT$ there are enough vertices $v\in \GG$ whose neighborhoods are similar to $\TT$, and that those vertices can be chosen to be far apart from each other and far from any small cycle. This axiom will be used in the elimination of the $\supl$ aggregator in Theorem~\ref{thm:aggelimsparse}. The supremum of a term is either obtained in the local neighborhood of the small cycles and roots (in the $r$-core), or somewhere in the remainder of the graph. Locally, this remainder part of the graph looks like a forest, and the richness axiom guarantees that for each finite-height tree $\TT$ there are enough trees that are similar to it. Here ``enough'' is modulated by the parameter $k$, and similarity is defined by $\sim_{k,\epsilon,\eta}$. Therefore, the supremum of the term on the remainder part is close to the supremum over all rooted trees, which is how we defined our controller in Subsection~\ref{subsec:linear_controllers}, where ``close'' is again modulated by the parameters $k, \epsilon, \eta, r$.

\begin{definition}[Richness axioms]
    We say that MRFG $\GG=(G,\bar v, \chi)$ satisfies the \emph{$(k,\epsilon, \eta ,r)$-richness axiom} (or simply ``is $(k,\epsilon, \eta ,r)$-rich'') if for any $r^\prime\leq r$ and any finite rooted featured tree $\TT$ whose height is at most $r^\prime$, there are $k$ vertices $v_1,\dots, v_k\in V(G)$ such that the following hold.
    \begin{enumerate}[label=(\arabic*)]
        \item
        $\ball_{r^\prime}(v_i) \sim_{k,\epsilon,\eta} \TT$        
        for all $i\in [k]$.
        \item For distinct $i,j\in [k]$ the distance between $v_i$ and $v_j$ is greater than $2r+1$.
        \item For all $i\in [k]$, the
        distance from $v_i$ to any root of $\GG$ or any cycle of length at most $2r +1 $ is greater than
        $2r + 1$.
    \end{enumerate}
\end{definition}

Next, the FBP axiom states that the neighborhood distribution of $\GG$ can be approximated by $\FBP$. This will be used in the elimination of the global $\Mean$ aggregator in Theorem~\ref{thm:aggelimsparse}. When considering the global mean of a term, we can again divide the graph into the $r$-core, and the remainder. As long as the first part is a small proportion of the whole graph, the global mean is close to the mean on the remainder. The FBP axiom states that this remainder looks locally like an $\FBP$ random forest, so the mean over it is close to the limit of the mean over this random forest. This matches how the global mean is handled in our controller definition, in Subsection~\ref{subsec:linear_controllers}. 

\begin{definition}[FBP axioms]
    We say that $\GG$ is \emph{$(k,\epsilon, \eta , r)$-close} to $\FBP$ if there is a coupling $\Pi$ of $V(G)$ and $\FBP\vert_r$ satisfying:\footnote{Recall that coupling with the set $V(G)$ means coupling with a uniform random variable over $V(G)$.}
    \begin{equation}
        \label{eq:close_coupling}
        \Pr_{(v,\TT)\sim \Pi}
        \left(
        \ball_{r}(v) \sim_{k,\epsilon,\eta} \TT
        \right) \geq 1 - \eta.
    \end{equation}
    We say that a MRFG satisfies the \emph{$(k,\epsilon, \eta, r)$-$\FBP$ axiom} (or just ``is $(k,\epsilon, \eta, r)$-similar to $\FBP$'') if it is $(k,\epsilon, \eta, r^\prime)$-close to $\FBP$
    for all $r^\prime \leq r$. 
\end{definition}

Finally, the following homogeneity axiom entails that the $r$-neighborhood of
all small cycles together with the $r$-neighborhood of any small set of vertices can only amount to a small proportion of the whole vertex set of $\GG$. This is used again in the global $\Mean$ aggregator elimination. The homogeneity axiom will guarantee that the $r$-core, which is the local neighborhood of small cycles and roots, is a small proportion of the graph. This is required for the application of the FBP axiom, as described above.  

\begin{definition}[Homogeneity axioms] 
    \label{def:hom} 
    We say that a MRFG $\GG$ satisfies the \emph{$(k,\eta, r)$-homogeneity axiom} (or simply ``is $(k,\eta, r)$-homogeneous'')
    if:
    \[
    \frac{(k+\mathrm{Cycle}_r(G))\Delta(G)^{r}}{|V(G)|} \leq \eta
    \]
    where $\mathrm{Cycle}_r(G)$ stands for the number of vertices in $G$ that belong to a cycle of length at most $2r + 1$ and we recall from the preliminaries that $\Delta(G)$ is the maximum degree in $G$.
\end{definition}

We note that the notions of richness, homogeneity, and similarity to $\FBP$ are preserved under expansion, and also preserved under decreasing the granularity of similarity used. See Appendix \ref{app:closureproperties} for a proof of the following lemma.

\begin{lemma}[Closure properties]
    \label{lem:inductive_axioms}
    Let $k,r \geq 0$ be integers, and let $\epsilon,\eta>0$.
    Let $\GG$ be a MRFG and $v\in V(G)$. The following hold:
    \begin{enumerate}[label=(\Roman*)]
        \item Suppose that $\GG$ is 
        $(k,\epsilon, \eta, 3r+1)$-rich. Then $\GG[v]$ is
        $(k-1, \epsilon, \eta, r)$-rich.
        \item Suppose that $\GG$ is $(k,\epsilon,\eta,r)$-similar to $\FBP$. Then $\GG[v]$ also has this property. 
        \item Suppose that $\GG$ is $(k,\eta,r)$-homogeneous.
        Then $\GG[u]$ also has this property.
        \item Suppose that $\GG$ is 
        $(k,\epsilon,\eta,r)$-rich. Then it is also $(k^\prime,\epsilon,\eta,r^\prime)$-rich for any $k^\prime\leq k$, $r^\prime \leq r$.
         \item Suppose that $\GG$ is $(k,\epsilon,\eta,r)$-similar to $\FBP$. Then it is also $(k^\prime, \epsilon, \eta, r^\prime)$-similar to $\FBP$ for any $k^\prime\leq k$, $r^\prime \leq r$.
        \item Suppose that $\GG$ is $(k,\eta,r)$-homogeneous. Then it is also $(k^\prime, \eta, r^\prime)$-homogeneous to $\FBP$ for any $k^\prime \leq k, r^\prime \leq r$.
    \end{enumerate}
\end{lemma}

We are now ready to give the model-theoretic result. Informally, this says that for any feature graphs satisfying the axioms we have defined, a term is close to its controller.

\begin{theorem}[Aggregate elimination when the axioms hold]
\label{thm:aggelimsparse}
     Let $K, k_1, k_2\geq 0$ be  integers satisfying $k_1+k_2=K$, and $\epsilon, C > 0$,
    $\eta= \frac{\epsilon}{4C}$.
    Let $\tau(\bar w)\in \Aggo{\Mean,\LMean,  \supl}$ be a term with $|\bar w|\leq k_1$, $\rank(\tau)\leq k_2$, 
    and $|\lambda_{\tau^\prime}| \leq C$ for all subterms $\tau^\prime$ of $\tau$.
    Let $\GG$ be an MRFG with $|\bar w|$ roots that is:
    \begin{itemize}
        \item $(k_2,\epsilon,\eta,r_{k_2})$-rich,
        \item $(k_2,\epsilon,\eta,r_{k_2})$-close to $\FBP_{r_{k_2}}$, and
        \item $(K,\eta, r_K)$-homogeneous,
    \end{itemize}
    where we define $r_i=\frac{3^i-1}{2}$ for all $i\geq 0$. Then:
    \[
         \left\vert \dsb{\tau}_{\GG} -
         \lambda_\tau(\GG)) \right\vert 
        \leq \epsilon \cdot \slope_\tau \cdot k_2.
    \]  
\end{theorem}

\begin{proof}
Fix $\epsilon, C>0, \eta=\frac{\epsilon}{4C}$ and $K\geq 0$. The proof is by induction on $k_2$. 
    For $k_2=0$, we have $\rank(\tau)=0$ and the statement is straightforward.
    For the inductive step, let $k_2>0$ and assume the statement holds for smaller values.
    
    \textbf{Supremum step: $\tau(\bar u)=\sup_v \pi(\bar u, v)$.}
    We need to compare
    $\dsb{\tau}_{\GG}$ to $\lambda_\tau(\GG)$. 
    Let $u\in V(G)$ be arbitrary. By the closure properties in Lemma \ref{lem:inductive_axioms}, $\GG[u]$ is still:
    \begin{itemize}
        \item $(k_2-1,\epsilon,\eta,r_{k_2-1})$-rich,
        \item $(k_2-1,\epsilon,\eta,r_{k_2-1})$-close to $\FBP_{r_{k_2-1}}$, and
        \item $(K,\eta, r_K)$-homogeneous.
    \end{itemize}
    Hence, we can apply the induction hypothesis to $\pi$, obtaining:
        \begin{equation}  \label{eq:model_main_aux1}
    \left\vert\dsb{\pi}_{\GG[u]}
        - \lambda_{\pi}(\GG[u])
        \right\vert \leq \epsilon \slope_\tau (k_2-1), 
    \end{equation}
    for all $u\in V(G)$.
    In particular, using the definition of $\lambda_\tau$, this shows that:
    \[
    \dsb{\tau}_{\GG} \leq \lambda_\tau(\GG) +   \epsilon \slope_\tau(k_2-1).
    \]
    Now we want to obtain the bound:
    \[
    \lambda_\tau(\GG)
    \leq   \dsb{\tau}_{\GG} +   \epsilon \slope_\tau k_2.
    \]
   Following the definition of $\lambda_\tau$, there are two sub-cases to consider to prove this inequality. 
    \begin{enumerate}[label=(\roman*)]
        \item There is some $u\in V(G)$ that satisfies:
        \[ 
        \lambda_\tau(\GG)=
        \lambda_{\pi}(\GG[u]).
        \]
        In this case 
        using \Cref{eq:model_main_aux1} we get:
        \[
        \lambda_\tau(\GG)
        \leq   \dsb{\tau}_{\GG} +   \epsilon \slope_\tau (k_2-1)
        \leq \dsb{\tau}_{\GG} +   \epsilon \slope_\tau k_2.
        \]
        \item
        It holds that:
        \begin{align*}
         \lambda_\tau(\GG)= 
        \sup_{\substack{ \TT \text{ finite rooted}\\ \text{featured tree}}} 
        \lambda_{\pi}(\GG\sqcup \TT).
        \end{align*}
        By the Core Determinacy Lemma, Lemma \ref{lem:controller_core}, it is enough to consider $\TT$ of height at most $r_{k_2-1}$.
        Let $\TT$ be any such tree.
        By the same lemma:
        \[
        \lambda_{\pi}(\GG\sqcup \TT)=
        \lambda_{\pi}(\HH_1),
        \]
        where $\HH_1= \GG\vert_{r_{k_2-1}} \sqcup \TT$.
        By the $(k_2,\epsilon,\eta, r_{k_2})$-richness assumption, we can find a vertex $u\in V(G)$ such that $\ball_{r_{k_2-1}}(u)$ is disjoint from 
        $\GG\vert_{r_{k_2-1}}$ and 
        $\ball_{r_{k_2-1}}(u)\sim_{k,\epsilon,\eta} \TT$.
        By the induction hypothesis:
        \[\lambda_{\pi}(\GG[u]) \leq 
        \dsb{\tau}_{\GG} +  \epsilon \slope_\tau (k_2 -1).
        \]        
        By the Core Determinacy Lemma, Lemma \ref{lem:controller_core}, it holds that:
        \[
        \lambda_{\pi}(\GG[u])= \lambda_{\pi}\left(\HH_2 \right),
        \]
        where $\HH_2=\GG\vert_{r_{k_2-1}} \sqcup 
        \ball_{r_{k_2-1}}(u)$.  On the other hand, the fact that
        $\ball_{r_{k_2-1}}(u)\sim_{k_2,\epsilon,\eta} \TT$ implies
        $\HH_1 \sim_{k_2-1,\epsilon,\eta}\HH_2$,
        so using the preservation of controllers by games, Theorem \ref{thm:controller_vs_controller_approx}, together with
        $\slope_\tau=\slope_\pi$ we obtain:
        \[
        \left\vert
        \lambda_{\pi}(\HH_1) -
        \lambda_{\pi}(\HH_2)
        \right\vert \leq 
         \epsilon \slope_\tau.
        \]
        Putting everything together we get:
        \[
        \lambda_{\pi}(\GG\sqcup \TT) =
                \lambda_{\pi}(\HH_1)
                \leq  \dsb{\tau}_{\GG} +  \epsilon \slope_\tau 2k_2,
        \]
        as we wanted to show.
    \end{enumerate}
    
        \textbf{Global mean step: $\tau(\bar u)=\Mean_y \pi(\bar u, v)$.}
        As before, we need to compare
    $\dsb{\tau}_{\GG}$ to $\lambda_\tau(\GG)$. 
    Let $u\in V(G)$ be arbitrary. Observe that $\GG[u]$ still satisfies the induction hypotheses. By definition:
    \[
    \lambda_\tau(\GG) =
    \Ex_{\TT\sim \FBP\vert_{r_{k_2 -1}}} 
    \left[  
    \lambda_{\pi}(\GG\sqcup \TT)
    \right],
    \]
    Similarly:
    \[
    \dsb{\tau}_{\GG} =
    \frac{1}{|V(G)|} \sum_{u\in V(G)}
    \dsb{\pi}_{\GG[u]}.
    \]
    By hypothesis, $\GG$ is $(k_2,\epsilon, \eta, r_{k_2})$-close to $\FBP\vert_{r_{k_2}}$, so there is a coupling $\Pi$ of $V(G)$ and $\FBP\vert_{r_{k_2}}$ satisfying \eqref{eq:close_coupling}. By the definition of $\dsb \tau_{\GG}$ and the triangle inequality we have that:
    \begin{align}
    \nonumber
        &\left\vert
        \lambda_\tau(\GG) -
        \dsb{\tau}_{\GG} 
        \right\vert
         \\
         \nonumber
    &\quad=\left\vert
    \Ex_{(u,\TT)\sim \Pi} \left[
        \lambda_{\pi}(\GG\sqcup \TT) -
        \dsb{\pi}_{\GG[u]} 
        \right]
        \right\vert \\    
    \label{eq:aux_mean0}
        &\quad\leq
        \left\vert
          \Ex_{(u,\TT)\sim \Pi}\left[
        \lambda_{\pi}(\GG\sqcup \TT) - \lambda_{\pi}(\GG[u])
        \right]
        \right\vert 
    \\ 
        \label{eq:aux_mean00}
    &\qquad+\left\vert
    \Ex_{(u,\TT)\sim \Pi}\left[
        \dsb{\pi}_{\GG[u]} -
        \lambda_{\pi}(\GG[u])
        \right]
    \right\vert. 
    \end{align}
We bound both terms separately. By the induction hypothesis, \eqref{eq:aux_mean00} is at most $\epsilon \cdot  \slope_\tau \cdot (k_2-1)$. Let us focus on \eqref{eq:aux_mean0}. \par
Let $V_{\mathrm{cl}}$ be the set consisting of all vertices $v\in V(G)$ that are at distance at most $2r_{k_2-1} + 1$ to some root in $\GG$
    or some cycle of size at most $2r_{k_2-1} +1$. That is, $V_\mathrm{cl}$ is the set of vertices that are ``close'' to $\GG\vert_{r_{k_2-1}}$. Using the fact that $\GG$ is $(k_2, \eta, r_K)$-homogeneous we obtain that $\frac{|V_\mathrm{cl}|}{n} \leq \eta= \frac{\epsilon}{4C}$. Indeed, the size of the $r_{k_2}$-core of $\GG$ is at most
    \[
    (K + \mathrm{Cycle}_{r_{k_2}}(G))\Delta(G)^{r_{k_2}},
    \]
    where $\mathrm{Cycle}_{r_{k_2}}$ denotes the number of vertices lying in some cycle of length at most $2r_{k_2}+1$ in $G$, as in the definition of the homogeneity axiom.
    Let $(u,\TT)\sim \Pi$, and let $A$ be the event that $\ball_{r_{k_2}}(u)\sim_{k_2, \epsilon,\eta} \TT$ and $u\notin V_\mathrm{cl}$. Using a union bound, we obtain that:
    \begin{equation}
    \label{eq:good_event}
    \Pr_{(u,\TT)\sim \Pi}((u,\TT)\in A) \geq 1 - \frac{\epsilon}{2C}.        
    \end{equation}
Considering the event $A$ and its negation $\neg A$ and using the triangle inequality, we obtain that \eqref{eq:aux_mean0} is at most:
\begin{align}
    &
    \left\vert
        \def\arraystretch{2}
        \begin{array}{l}
            \displaystyle
            \Pr_{(u,\TT)\sim \Pi}((u,\TT)\in A) \\
            \displaystyle
            \times 
            \Ex_{(u,\TT)\sim (\Pi\vert A)}
                \left[
            \lambda_{\pi}(\GG\sqcup \TT) - \lambda_{\pi}(\GG[u])
            \right]
        \end{array}
    \right\vert
    \label{eq:aux_mean1} \\
    & +
    \left\vert
        \def\arraystretch{2}
        \begin{array}{l}
            \displaystyle
            \Pr_{(u,\TT)\sim \Pi}((u,\TT)\not\in A) \\
            \displaystyle
             \times 
             \Ex_{(u,\TT)\sim (\Pi\vert \neg A)}
            \left[
                \lambda_{\pi}(\GG\sqcup \TT) - \lambda_{\pi}(\GG[u])
            \right]
        \end{array}
    \right\vert.
     \label{eq:aux_mean2}
\end{align}
We bound each term of this sum. Using \eqref{eq:good_event} and the fact that $|\lambda_{\pi}|\leq C$, we obtain that \eqref{eq:aux_mean2} is at most $\epsilon$. Let us consider \eqref{eq:aux_mean1}. Let $(u,\TT)\in A$. Then we claim that:
\begin{equation}
\label{eq:good_event_game}
\abs{\lambda_{\pi}(\GG\sqcup \TT) - \lambda_{\pi}(\GG[u])} \leq
\epsilon \cdot \slope_{\pi} .    
\end{equation}
Indeed, applying the Core Determinacy Lemma, Lemma \ref{lem:controller_core}, we get 
$\lambda_{\pi}(\GG\sqcup \TT)= \lambda_{\pi}((\GG\sqcup \TT)\vert_{r_{k_2}})$,
and $\lambda_{\pi}(\GG[u])= \lambda_{\pi}((\GG[u])\vert_{r_{k_2}})$.
As $\TT$ has height at most $r_{k_2}$, it holds that $(\GG\sqcup \TT)\vert_{r_{k_2}} 
= \GG\vert_{r_{k_2}} \sqcup \TT$. Similarly, using the fact that $u\notin V_\mathrm{cl}$,
we obtain $\GG[u]\vert_{r_{k_2}} = \GG\vert_{r_{k_2}} \sqcup \ball_{r_{k_2}}(u)$. The fact that
$\ball_{r_{k_2}}(u)\sim_{k_2,\epsilon,\eta} \TT$ implies that:
\begin{equation*}
    \left(
        \GG\vert_{r_{k_2}} \sqcup \TT
    \right)
    \sim_{k_2,\epsilon,\eta} 
    \left(
        \GG\vert_{r_{k_2}} \sqcup \ball_{r_{k_2}}(u)
    \right).
\end{equation*}
Hence, \eqref{eq:good_event_game} follows now from preservation of controllers by games, Theorem \ref{thm:controller_vs_controller_approx}.
This implies that \eqref{eq:aux_mean1} is at most $\epsilon \cdot \slope_{\pi}$. Thus, $\eqref{eq:aux_mean0} \leq \eqref{eq:aux_mean1} + \eqref{eq:aux_mean2} \leq \epsilon (\slope_{\pi}+ 1)= 
\epsilon \slope_\tau$. Finally:
\[
 \left\vert
        \lambda_\tau(\GG) -
        \dsb{\tau}_{\GG} 
        \right\vert \leq 
        \eqref{eq:aux_mean0} + \eqref{eq:aux_mean00} 
        \leq \epsilon \slope_\tau + \epsilon \slope_\tau(k_2-1) = 
        \epsilon \slope_\tau k_2,
\]
as we wanted to show. 

\textbf{Local mean step: $\tau\equiv \LMean_{v E u_i} \pi(\bar u, v)$.}
Let $u_i$ denote $\GG$'s $i^{th}$ root. We consider two subcases. If $u_i$ is isolated, then $\dsb{\tau}_\GG= \lambda_\tau(\GG)=0$. Otherwise, using the definitions of $\dsb{\tau}$
        and $\lambda_\tau$ we obtain:
        \begin{align*}
        \abs{\dsb{\tau}_\GG - \lambda_\tau(\GG)} \leq
        \frac{1}{|\Ne(u_i)|}
        \sum_{v \in \Ne(u_i)}
        \abs*{
        \dsb{\pi}_{\GG[v]} -
        \lambda_{\pi}(\GG[v])
        }
        \end{align*}
        By the induction hypothesis $ \abs{
        \dsb{\pi}_{\GG[v]} -
        \lambda_{\pi}(\GG[v])
        } \leq \epsilon \slope_\tau k_2$ for any choice of $v$, so
        $\abs{\dsb{\tau}_\GG - \lambda_\tau(\GG)} \leq \epsilon \slope_\tau k_2$, as we wanted to show. 

\textbf{Function application step: $\tau \equiv F(\pi_1,\dots, \pi_m)$.} Here we assume that the statement holds for $\pi_1,\dots, \pi_m$ by induction. Then, by this assumption:
\begin{align*}
    &\left\vert
       \dsb{\tau}_{\GG} - \lambda_\tau(\GG) 
    \right\vert \\
    &\quad= \left\vert
       F(\dsb{\pi_1}_{\GG},\dots, 
       \dsb{\pi_m}_{\GG}) - F(\lambda_{\pi_1}(\GG),\dots,  \lambda_{\pi_m}(\GG))
    \right\vert \\
    &\quad\leq 
    \slope_F \cdot \max_{i\in [m]}
    \left\vert
    \dsb{\tau_i}_{\GG} - \lambda_{\tau_i}(\GG)
    \right\vert \\
    &\quad\leq  \epsilon \cdot \slope_\tau \cdot  k_2.
\end{align*}

The last inequality uses the inductive definition of the slope $\slope_{\tau}$ for Lipschitz function terms $\tau$.
\end{proof}

\subsection{Combinatorial Part: The Axioms Hold on Most Graphs} \label{subsec:combinatorial}

The other ingredients of our proof involve showing properties about linear sparse random featured graphs that do not involve our term language.
We defer the proofs, which are variations of results known in the absence of features, to the appendix.

We need that the axioms hold on most graphs: see Appendices \ref{app:richnessholds}, \ref{app:bpholds}, and \ref{app:homogeneity}.

\begin{theorem}
    \label{thm:axioms hold aas}
    Let $k,r\geq 0$ be integers, and $\epsilon,\eta>0$. Then a.a.s.\@ $\Gcal_\Dcal(n,c/n)$ is:
    \begin{enumerate}[label=(\Roman*)]
        \item \label{item:richness; thm:axioms hold aas}
            $(k,\epsilon,\eta,r)$-rich,
        \item \label{item:fpb; thm:axioms hold aas}
            $(k,\epsilon, \eta, r)$-similar to FBP, and
        \item \label{item:homogeneity; thm:axioms hold aas}
            $(k, \eta, r)$-homogeneous.
    \end{enumerate}
\end{theorem}

We know that the axioms hold a.a.s. by the result above. Since controllers have bounded image (Proposition \ref{prop:diam_controller}), we can get an appropriate $C$ to apply
the result on aggregate elimination when the axioms hold 
(Theorem \ref{thm:aggelimsparse}) to  conclude that aggregate elimination holds asymptotically almost surely. Thus we have:

\begin{corollary}
\label{cor:convergence_in_probability}
    Let $\tau\in \Aggo{\Mean,\LMean, \supl}$ be a closed term. Then $\dsb\tau_{\Gcal_\Dcal(n,c/n)}$ converges in probability to $\lambda_\tau(\Gcal_\Dcal(n,c/n))$.
\end{corollary}

\subsection{Proof of the main result, Theorem \ref{thm:mainthmlinearsparse}}
\label{subsec:linear_main_proof}

Let $\tau\in \Aggo{\Mean,\LMean,\supl}$ be a closed term, let $k=\rank(\tau)$ and let $r_k=\frac{3^k-1}{2}$. Recall the definition of the random core $\mathrm{Core}_{r_k}$, and the random featured core $\mathrm{Core}_{r_k,\Dcal}$ from Subsection \ref{subsec:linear_prelims}. We show that $\tau$ converges in distribution to $\lambda_\tau(\mathrm{Core}_{r_k,\Dcal})$.
By the a.a.s. simplification result, Corollary \ref{cor:convergence_in_probability}, it is enough to show that 
$\lambda_\tau(\Gcal_\Dcal(n,c/n))$ converges in distribution to $\lambda_\tau(\mathrm{Core}_{r_k,\Dcal})$. Indeed, suppose that this is the case, and let $x$ be a continuity point of the map $y\mapsto \Pr(\lambda_\tau(\mathrm{Core}_{r_k,\Dcal}) \leq y)$. By Corollary \ref{cor:convergence_in_probability}, for every $\delta\geq 0$ it holds that:
\begin{align*}
&\lim_{n\to\infty} \Pr( \lambda_\tau(\Gcal_\Dcal(n,c/n)) \leq x- \delta ) \\ 
&\quad\leq
\lim_{n\to\infty} \Pr( \dsb{\tau}_{\Gcal_\Dcal(n,c/n)} \leq x ) \\
&\quad\leq
\lim_{n\to\infty} \Pr( \lambda_\tau(\Gcal_\Dcal(n,c/n)) \leq x + \delta ).  
\end{align*}
However, by assumption we have that the function defined as $y\mapsto  \Pr( \lambda_\tau(\Gcal_\Dcal(n,c/n)) \leq y)$ converges pointwise to the function
$y\mapsto \Pr(\lambda_\tau(\mathrm{Core}_{r_k,\Dcal}) \leq y)$, which is 
continuous at $x$. Hence:
\[
\lim_{n\to\infty} \Pr( \dsb{\tau}_{\Gcal_\Dcal(n,c/n)} \leq x ) =
\Pr(\lambda_\tau(\mathrm{Core}_{r_k,\Dcal}) \leq x),
\]
as we wanted to prove.\par
We move on to proving that $\lambda_\tau(\Gcal_\Dcal(n,c/n))$ converges in distribution to $\lambda_\tau(\mathrm{Core}_{r_k,\Dcal})$.
For each $n\geq 0$, let $\mathbb{K}_n=(K_n, \chi_n)$ be the random featured graph $\Gcal_\Dcal(n,c/n)|_{r_k}$. We again apply Core Determinacy, Lemma \ref{lem:controller_core},  to get $\lambda_\tau(\Gcal_\Dcal(n,c/n))= 
\lambda_\tau(\mathbb{K}_n)
$. We show that for all $y\in \mathbb{R}$:
\begin{equation}
\label{eq:main_1}
\lim_{n\to\infty}\Pr(\lambda_\tau(\mathbb{K}_n) \leq y) =
\Pr(\lambda_\tau(\mathrm{Core}_{r_k,\Dcal}) \leq y). 
\end{equation}
For this it is enough to prove that for all $\nu>0$:
\[
\left| \lim_{n\to\infty}\Pr(\lambda_\tau(\mathbb{K}_n) \leq y) -
\Pr(\lambda_\tau(\mathrm{Core}_{r_k,\Dcal}) \leq y)\right| < \nu.
\] 
Let $H_1,\dots, H_\ell$ be a family of graphs satisfying that:
\[
\Pr\left(\bigwedge_{i=1}^\ell \mathrm{Core}_{r_k} \not\simeq H_i\right) < \nu/3
\] 
Let $p_i= \Pr(\mathrm{Core}_{r_k}\simeq H_i)$ 
for each $i=1,\dots \ell$. 
Observe that given a fixed graph $H$, the distribution of $\mathbb{K}_n$ conditioned on $K_n\simeq H$
is the same as the distribution of $\mathrm{Core}_{r_k,\Dcal}$, 
conditioned on its underlying graph,
$\mathrm{Core}_{r_k}$, being isomorphic to $H$. 
Indeed, in both cases the underlying graph is fixed and equal, and the features are distributed independently according to $\Dcal$.\par
Using that $\lim_{n\to \infty} \Pr(K_n\simeq H_i) = p_i$ by Fact \ref{fact:core_tight_distribution},
we obtain:
\begin{align*}
    \left|
        \def\arraystretch{2}
        \begin{array}{l}
            \displaystyle
            \lim_{n\to\infty} \Pr(\lambda_\tau(\mathbb{K}_n) \leq y) \\
            \displaystyle
            - \sum_{i=1}^\ell
            p_i \Pr(\lambda_\tau(\mathrm{Core}_{r_k,\Dcal}) \leq y \, \big\vert \, \mathrm{Core}_{r_k}\simeq H_i) 
        \end{array}
    \right|
    \leq \frac{\nu}{3}.
\end{align*}

Using this inequality it follows that:
\begin{align*}
    & \left| 
        \lim_{n\to\infty}\Pr(\lambda_\tau(\mathbb{K}_n) \leq y) -
        \Pr(\lambda_\tau(\mathrm{Core}_{r_k,\Dcal}) \leq y)
    \right|  \\ 
    &\quad\leq
    \left|
        \def\arraystretch{2}
        \begin{array}{l}
            \displaystyle
            \lim_{n\to\infty}\Pr(\lambda_\tau(\mathbb{K}_n) \leq y) \\
            \displaystyle
            - 
            \sum_{i=1}^\ell
            p_i \Pr(\lambda_\tau(\mathrm{Core}_{r_k,\Dcal}) \leq y \, \big\vert \,  \mathrm{Core}_{r_k}\simeq H_i)
        \end{array}
    \right| \\
    &\qquad +
    \left|
        \def\arraystretch{2}
        \begin{array}{l}
            \displaystyle
            \Pr(\lambda_\tau(\mathrm{Core}_{r,\Dcal}) \leq y) \\
            \displaystyle
            - 
            \sum_{i=1}^\ell 
            p_i \Pr(\lambda_\tau(\mathrm{Core}_{r_k,\Dcal}) \leq y \, \big\vert \,  \mathrm{Core}_{r_k}\simeq H_i)
        \end{array}
    \right| \\
    &\quad \leq 2\nu/3 
    < \nu.
\end{align*}
This proves~(\ref{eq:main_1}), and finishes the proof of Theorem \ref{thm:mainthmlinearsparse}. 
\qed

\section{Discussion} \label{sec:conc} 
Our paper presents  convergence laws for a real-valued logic extending first-order logic with averaging operators.

We do not discuss computational issues here, and clearly computing the limit values of terms requires us to restrict the term language, which allows arbitrary Lipschitz functions.  Under reasonable assumptions on the functions, we believe that a PSPACE
bound can be easily extracted for computing the a.a.s. probability in the dense case --- our controller-based algorithm can be seen as an extension of the PSPACE algorithm for first-order logic from \cite{grandjean}. For the linear sparse case, \cite{lynchlinearsparse} has obtained expressions for the probabilities of first-order logic terms, but we have not considered how to extend this approach to compute limit probabilities for our term language.

Let us now discuss the status of convergence laws for other term languages and other distributions.

We start with the question of other operators.
Here, we focused on local and global averaging on featured graphs, but we believe that our results also hold for other average-based operators, like the weighted average operator considered in \cite{usneurips24}: the tools we developed, particularly the games for averaging, were constructed with such a generalization in mind. We also believe that our results can generalize to arbitrary arity relational structures, but have not investigated what such a generalization would look like.
Another interesting question is what happens for summation-based aggregation. While one clearly cannot get convergence for general terms, it is possible that one can characterize subsequences of $\N$ on which one converges, in the spirit of results for logics with parity quantification \cite{kolaitisparity}.

We turn now to other distributions.
A generalization that we believe not to be difficult, is the \emph{sublinear sparse case} of {\erdosrenyi}, where the edge probability is $O(\frac{1}{n^\beta})$ for $\beta >1$. In the first-order setting, a zero-one law holds when $\frac{\ell+1}{\ell}<\beta<\frac{\ell+2}{\ell+1}$ for some integer $\ell\geq 1$,
or when $\beta>2$ \cite{shelahspencersparse}, and a convergence law holds when $\beta= \frac{\ell+1}{\ell}$ for some integer $\ell\geq 1$ \cite{lynchlinearsparse}. The more interesting case is
where $\beta=\frac{\ell+1}{\ell}$.  Here a.a.s. almost all vertices are isolated, which simplifies the analysis of the global mean operator. Asymptotically almost surely there are no cycles and no components containing more than $\ell+1$ vertices.
For any tree $T$ containing at most $\ell$ vertices, a.a.s. the number of components isomorphic to $T$ is unbounded. This leads to a simpler version of our richness axiom. In this setting the part of the graph that determines the value of first-order sentences a.a.s. is the union of all components containing $\ell+1$ vertices. We believe that the same approach works for our aggregate term language, and will yield a convergence law.

There are two cases of \erdosrenyi\ that are more challenging.
One is \emph{logarithmic growth}: \cite{spencerthoma} showed convergence for first-order logic for the case of growth  $\frac{\log(n)}{n}$.
We believe that a similar analysis to what we present here would allow us to obtain a convergence law for this case, but we have not verified this.
We also leave open the case of $n^{-\alpha}$ for $\alpha$ irrational: a convergence result here would require an extension of the intricate argument due to Shelah and Spencer for first-order logic \cite{shelahspencersparse}.

Note that almost sure convergence for first-order logic and convergence
for averaging operators alone --- our term language with $\supl$ removed --- are incomparable: in rational root growth cases, like $\frac{1}{\sqrt{n}}$, we know that averaging operators have strong convergence \cite{usneurips24}, while first-order logic does not have any convergence \cite{shelahspencersparse}. On the other hand, if we consider $p(n)=\frac{1}{2}$ for $n$ even and $\frac 1 3$ for $n$ odd, first-order logic has a zero-one law. But it is easy to see that averages will converge to a different value on the evens and the odds.

As noted in the related work section, there are numerous convergence results outside of the context of {\erdosrenyi}: for example, for uniform distributions over sparse graph classes \cite{lynchdegreesequence,kolaitisfreegraphs,noyminorclosed}. Our work leaves open the possibility that these extend to aggregate logics, but we do not investigate this here.

\bibliographystyle{alpha}
\bibliography{main}

\newpage
\onecolumn
\appendix

\section{Hoeffding's Inequality}\label{app:hoeffding}

\begin{theorem}[Hoeffding's Inequality for bounded random variables]
    \label{thm:Hoeffding's Inequality for bounded random variables}
    Let $X_i$ for $i \leq n$ be i.i.d.\@ bounded random variables taking values in $[a,
    b]$ with common mean $\mu$. Then for any $\lam > 0$ we have:
    \begin{equation*}
        \Pr\left(\abs*{\sum_{i=1}^n X_i - n\mu} \geq \lam\right)
        \leq
        2 \exp\left(-\frac {2\lam^2} {n(b-a)^2}\right)
    \end{equation*}
\end{theorem}

\begin{proof}
    See Theorem~2.2.6, p.~16 of \cite{vershynin}.
\end{proof}

\begin{corollary}[Hoeffding's Inequality for Bernoulli random variables]
    \label{cor:Hoeffding's Inequality for Bernoulli random variables}
    Let $X_i$ for $i \leq n$ be i.i.d.\@ Bernoulli random variables with parameter $p$.
    Then for any $\lam > 0$ we have:
    \begin{equation*}
        \Pr\left(\abs*{\sum_{i=1}^n X_i - np} \geq \lam\right)
        \leq
        2 \exp\left(-\frac {2\lam^2} n\right)
    \end{equation*}
\end{corollary}
\section{Distance from Supremum in the Dense Random Graph Model} \label{app:supbound}

Recall Lemma \ref{lem:probability of being close to sup}:

\medskip

    Let $X, Y$ be compact Euclidean domains, take $f \colon X \times Y \to \R$ Lipschitz continuous, and let $\mathcal C$ be a distribution with support $Y$. Then for every $\margin > 0$ we have that:
    \begin{equation*}
        \inf_{x \in X}
        \Pr_{y \sim \mathcal C}\left(
            f(x, y) \geq \sup_{y' \in Y} f(x, y') - \margin
        \right)
        > 0
    \end{equation*}

\medskip

\begin{proof}
    Define $q \colon X \to \R$ by:
    \begin{equation*}
        q(x)
        \coloneqq
        \Pr_{y \sim \mathcal C}\left(
            f(x, y) \geq \sup_{y' \in Y} f(x, y') - \margin
        \right)
    \end{equation*}
    First, for any fixed $x \in X$, we can define the real-valued random variable $Z_x$ obtained by sampling $y \sim \mathcal C$ and computing $f(x, y)$. Then $Z_x$ has support $f[\{x\} \times Y]$, which is compact, and:
    \begin{equation*}
        q(x) = \Pr(Z_x \geq \sup f[\{x\} \times Y] - \margin) > 0
    \end{equation*}
    Because $X$ is compact, to prove the result it suffices to show that $q$ is continuous.

    Consider a sequence $(x_n)$ in $X$ converging to $x$. Let:
    \begin{equation*}
        A_n \coloneqq \left\{y \in Y \;\middle|\; f(x_n, y) \geq \sup_{y' \in Y} f(x_n, y') - \margin\right\}
    \end{equation*}
    and:
    \begin{equation*}
        A \coloneqq
        \left\{y \in Y \;\middle|\; f(x, y) \geq \sup_{y' \in Y} f(x, y') - \margin\right\}
    \end{equation*}
    It is not hard to see, using the Lipschitz continuity of $f$, that:
    \begin{equation*}
        \limsup A_n 
        = \liminf A_n
        = A
    \end{equation*}
    Therefore, by the continuity of probability, we have that:
    \begin{equation*}
        \lim_{n \to \infty} q(x_n) 
        = \lim_{n \to \infty} \Pr(A_n)
        = \Pr(A)
        = q(x)
    \end{equation*}
\end{proof}
\section{Controllers have Bounded Images: Proof of Proposition \ref{prop:diam_controller}} \label{app:controllersboundedimage}

Recall the statement of
Proposition \ref{prop:diam_controller}:

\medskip

    Let $\tau(\bar u) \in \Aggo{\Mean, \LMean, \supl}$ be a term. Then $\abs{\lambda_\tau}$ is bounded.

\medskip

\begin{proof}
The statement is proven inductively on terms. If $\tau(\bar u)$ is an atomic term, then it satisfies the statement. If $\tau\equiv F(\pi_1,\dots, \pi_m)$ for some Lipschitz function $F$ and terms $\pi_1, \dots \pi_m$ satisfying that $|\lambda_{\pi_i}|$ is bounded for each $i\in [m]$, then $|\lambda_\tau|$ is also bounded. If $\tau(\bar u) \equiv \sup_v \pi(\bar u, v)$, then 
$\mathrm{Im}(\lambda_\tau) \subseteq \mathrm{Im}(\lambda_{\pi})$. Hence, if $\pi$ satisfies the statement, so does $\tau$. If $\tau(\bar u) \equiv \sup_v \pi(\bar u, v)$, then $\mathrm{Im}(\lambda_\tau)$ is contained in the convex hull of $\mathrm{Im}(\lambda_{\pi})$. Hence, as before, if $\pi$ satisfies the statement, so does $\tau$. Finally, suppose that
$\tau(\bar u, v)= \LMean_{v E w} \pi(\bar u, v, w)$, and let $\GG$ be a MFRG with $|\concat{\bar u}{v}|$ roots. Then 
$\lambda_{\tau}(\GG)$ is either zero (if the last root of $\GG$ is an isolated vertex) or belongs to the convex hull of $\mathrm{Im}(\lambda_{\pi})$. Hence $\tau$ satisfies the statement. 
\end{proof}

\section{Core Determinacy of Controllers: Proof of Lemma \ref{lem:controller_core}} \label{app:coredeterminacy}

Recall Lemma
\ref{lem:controller_core}:

\medskip

Let $\tau(\bar u)\in \Aggo{\Mean, \LMean, \supl}$ be a term, and let $k=\srank(\tau) + \lmrank(\tau)$. Then, for each MRFG $\GG$ we have that:
\[
\lambda_\tau(\GG)=
\lambda_\tau(\GG\vert_{r_k}).
\]

\medskip

We will use the following simple results about cores and union:

\begin{obs}
\label{obs:core_aux}
    Let $\GG$, $\HH$ be MRFGs and $r\geq r^\prime \geq 0$ be integers. The following facts hold:
    \begin{itemize}
        \item $\coreof{\GG \sqcup \HH}{r} = \coreof{\GG}{r} \sqcup \coreof{\HH}{r}$.
        \item $(\GG\vert_r)\vert_{r^\prime} = \GG\vert_{r^\prime}$.
        \item Let $v\in V(G)$ be such that $\ball_{r^\prime}(v) \subseteq \GG\vert_r[v]$. Then $(\GG\vert_r[v])\vert_{r^\prime} = 
        \GG[v]\vert_{r^\prime}$.
    \end{itemize}
\end{obs}

\begin{proof}[Proof of Lemma \ref{lem:controller_core}]
    We prove the statement inductively on the structure of $\tau$. If $\tau$ is atomic, it clearly satisfies the statement. 
    The same is true if $\tau \equiv F(\pi_1,\dots, \pi_m)$ for some Lipschitz function $F$ and terms
    $\pi_1,\dots, \pi_m$,
    and the statement is assumed to hold for $\pi_1,\dots, \pi_m$. Suppose that $\tau(\bar u)
    \equiv \Mean_v \pi(\bar u, v)$, and $\pi$ satisfies the statement. Let $\GG$ be a 
    MRFG with $|\bar u|$ roots. Then:
    \[
    \lambda_\tau(\GG) = \Ex_{\TT\sim \FBP}\left[ 
    \lambda_{\pi}(\GG \sqcup \TT)
    \right].
    \]
    By the induction hypothesis:
    \[\lambda_{\pi}(\GG \sqcup \TT)=\lambda_{\pi}((\GG \sqcup \TT)\vert_{r_k}) 
    = \lambda_{\pi}(\GG\vert_{r_k} \sqcup \TT\vert_{r_k}).
    \]
    Here we have used Observation \ref{obs:core_aux} for the second equality. Now we argue that $\lambda(\GG\vert_{r_k})$ also equals this quantity. Indeed:
    \begin{align*}    
    \lambda_{\pi}(\GG\vert_{r_k} \sqcup \TT)=\lambda_{\pi}((\GG\vert_{r_k} \sqcup \TT)\vert_{r_k}) 
    &= \lambda_{\pi}((\GG\vert_{r_k})\vert_{r_k} \sqcup \TT\vert_{r_k}) \\
    &= \lambda_{\pi}(\GG\vert_{r_k} \sqcup \TT\vert_{r_k}).
    \end{align*}
    Here we have used Observation \ref{obs:core_aux} in the second and third equalities. \par
    Suppose that $\tau(\bar u) \equiv \LMean_{v E u_i} \pi(\bar u, v)$, where $\pi$ satisfies the statement, and 
    let $\GG$ be an arbitrary MRFG with $|\bar u|$ roots. Let $v\in V(G)$ be $\GG$'s $i$-th root. We have two cases. Suppose that $v$ is isolated. Then $\lambda_\tau(\GG)= \lambda_\tau(\GG\vert_{r_k})=0$. Otherwise, by induction:
    \begin{align}
    \nonumber
    \lambda_\tau(\GG) &= \frac{1}{|\Ne(v)|} \sum_{u\in \Ne(v)} \lambda_{\pi}(\GG[u])
    \\ 
    \label{eq:core_det1}
    &=\frac{1}{|\Ne(v)|} \sum_{u\in \Ne(v)} \lambda_{\pi}(\GG[u]\vert_{r_{k-1}}).
    \end{align}
    On the other hand:
    \begin{align}
    \nonumber
    \lambda_\tau(\GG\vert_{r_k})
    &= \frac{1}{|\Ne(v)|} \sum_{u\in \Ne(v)} \lambda_{\pi}(\GG\vert_{r_k}[u]) \\ 
    \label{eq:core_det2}
    &=\frac{1}{|\Ne(v)|} \sum_{u\in \Ne(v)} \lambda_{\pi}((\GG\vert_{r_k}[u])\vert_{r_{k-1}}).
    \end{align}
    Let $u\in \Ne(v)$. Using the fact that $u$ is adjacent to one of $\GG$'s roots and $r_k \geq r_{k-1} +1$ we obtain that 
    $\ball_{r_{k-1}}(u)\subseteq \GG\vert_{r_k}[u]$. Thus, by Observation \ref{obs:core_aux} $(\GG\vert_{r_k}[u]) \vert_{r_{k-1}} = (\GG[u]) \vert_{r_{k-1}}$. This proves that $\eqref{eq:core_det1}=\eqref{eq:core_det2}$, so 
    $\lambda_\tau(\GG)=\lambda_\tau(\GG\vert_{r_k})$.
    \par
    Finally, suppose that
    $\tau(\bar u)\equiv \sup_v \pi(\bar u, v)$,
    and $\pi$ satisfies the statement. Let $\GG$ be an arbitrary MRFG with $\abs{\bar u}$ roots. 
    We prove that $\lambda_\tau(\GG)\leq \lambda_\tau(\GG\vert_{r_k})$. The reverse inequality can be shown in a similar way.
    There are three cases.
    \begin{enumerate}[label=(\roman*)]
        \item \label{item:sup on tree; lem:controller_core}
        Suppose that:
        \begin{equation}
        \label{eq:controller_core_aux1}
        \lambda_\tau(\GG)= 
            \sup_{\substack{ \TT \text{ rooted}\\ \text{featured tree}}} 
            \lambda_{\pi}(\GG\sqcup \TT).
        \end{equation}      
        Let $\TT$ be a finite rooted featured tree $\TT$.
        By the induction assumption:
        \[\lambda_{\pi}(\GG\sqcup \TT)
        = \lambda_{\pi}((\GG\sqcup \TT)\vert_{r_{k-1}})
        = \lambda_{\pi}((\GG\vert_{r_{k-1}} \sqcup \TT\vert_{r_{k-1}}).
        \]
        Here the second equality follows from Observation \ref{obs:core_aux}. 
        Hence, we can rewrite \eqref{eq:controller_core_aux1} as:
        \[
            \lambda_\tau(\GG)= 
            \sup_{\substack{ \TT \text{rooted}\\ \text{featured tree}}} 
            \lambda_{\pi}(\GG\vert_{r_{k-1}}\sqcup \TT\vert_{r_{k-1}}).
        \]
        Arguing as in the $\Mean$ case, we obtain that
        $\lambda_{\pi}(\GG\vert_{r_k}\sqcup \TT) =
        \lambda_{\pi}((\GG\vert_{r_{k-1}} \sqcup \TT\vert_{r_{k-1}})$. 
        By the definition of $\lambda_\tau$, this implies that
        $\lambda_\tau(\GG\vert_{r_k})\geq \lambda_{\pi}((\GG\vert_{r_{k-1}} \sqcup \TT\vert_{r_{k-1}})$. Hence:
        \begin{gather*}
            \lambda_\tau(\GG) = 
            \sup_{\substack{ \TT \text{ rooted}\\ \text{featured tree}}} 
            \lambda_{\pi}(\GG\vert_{r_{k-1}}\sqcup \TT\vert_{r_{k-1}}) 
            \leq \lambda_\tau(\GG\vert_{r_k}),
        \end{gather*}
        as we wanted to show.
        \item Suppose that $\lambda_\tau(\GG)= \lambda_{\pi}(\GG[u])$ for some vertex $u\in V(G)$ that is at
        distance at most $2r_{k-1}$ 
        from some root of $\GG$ or some
        cycle of length at most $2r_{k-1} + 1$.
        We show that:
        \[\lambda_{\pi}(\GG[u])
        = \lambda_{\pi}(\GG\vert_{r_k}[u]) \leq
        \lambda_\tau(\GG\vert_{r_k}).
        \]
        The last inequality follows from the controller definition. We prove the first identity. By induction:
        \[
        \lambda_{\pi}(\GG[u]) =
        \lambda_{\pi}(\GG[u]\vert_{r_{k-1}}).
        \]
        Using that
        $r_k = 3r_{k-1} + 1$ we obtain  $\ball_{r_{k-1}}(u)\subseteq \GG\vert_{r_{k}}[u]$, so by Observation \ref{obs:core_aux} 
        $(\GG\vert_{r_{k}}[u])\vert_{r_{k-1}} = \GG[u]\vert_{r_{k-1}}$.
        Hence:
        \begin{align*}
        \lambda_{\pi}(\GG[u]) =
        \lambda_{\pi}(\GG[u]\vert_{r_{k-1}})
        &= \lambda_{\pi}((\GG\vert_{r_k}[u])\vert_{r_{k-1}}) \\
        &= \lambda_{\pi}(\GG\vert_{r_k}[u]),
        \end{align*}
        as we wanted to show. Here the second and third identities use the induction hypothesis.
        \item Suppose that $\lambda_\tau(\GG)= \lambda_{\pi}(\GG[u])$ for some vertex $u\in V(G)$ that is at distance greater than $2r_{k-1}$ to all roots in $\bar v$ and all cycles of length at most $2r_{k-1} + 1$. Then $\GG[u]\vert_{r_{k-1}}$ is the disjoint union of $\GG\vert_{r_{k-1}}$ and the featured rooted tree $\TT=\ball_{r_{k-1}}(u)$. This way:
        \begin{align*}
        \lambda_\tau(\GG)= \lambda_{\pi}(\GG[u]) 
        &= \lambda_{\pi}(\GG\vert_{r_{k-1}} \sqcup 
        \TT\vert_{r_{k-1}}) \\
        &= \lambda_{\pi}(\GG\sqcup\TT).
        \end{align*} 
        Hence, this is a particular case of scenario \ref{item:sup on tree; lem:controller_core} above.\qedhere
    \end{enumerate}
\end{proof}

\section{Proof of closure properties, Lemma \ref{lem:inductive_axioms}} \label{app:closureproperties}

Recall Lemma \ref{lem:inductive_axioms} from the body:

\medskip

    Let $k,r \geq 0$ be integers, and let $\epsilon,\eta>0$.
    Let $\GG$ be a MRFG and $v\in V(G)$. The following hold:
    \begin{enumerate}
        \item Suppose that $\GG$ is 
        $(k,\epsilon, \eta , 3r+1)$-rich. Then $\GG[v]$ is
        $(k-1, \epsilon, \eta, r)$-rich.
        \item Suppose that $\GG$ is $(k,\epsilon, \eta ,r)$-similar to $\FBP$ axiom. Then $\GG[v]$ also has this property. 
        \item Suppose that $\GG$ is $(k,\eta,r)$-homogeneous.
        Then $\GG[u]$ also has this property.
        \item Suppose that $\GG$ is 
        $(k,\epsilon,\eta ,r)$-rich. Then it is also $(k^\prime,\epsilon,\eta,r^\prime)$-rich for any $k^\prime\leq k$, $r^\prime \leq r$.
         \item Suppose that $\GG$ is $(k,\epsilon,\eta ,r)$-similar to $\FBP$ axiom. Then it is also $(k^\prime, \epsilon, \eta, r^\prime)$-similar to $\FBP$ for any $k^\prime\leq k$, $r^\prime \leq r$.
        \item Suppose that $\GG$ is $(k,\eta,r)$-homogeneous. Then it is also $(k^\prime, \eta, r^\prime)$-homogeneous to $\FBP$ for any $k^\prime \leq k, r^\prime \leq r$.
    \end{enumerate}

\medskip

\begin{proof}
The last three items follow from unrolling the axiom definitions, considering the fact that the $\sim_{k,\epsilon,\eta}$ relation refines
$\sim_{k^\prime, \epsilon,\eta}$ for any $k^\prime \leq k$. Items (2) and (3)
are also straightforward: neither similarity to $\FBP$ or 
homogeneity depend on the roots of a given MFRG. \par 
Let us show the first item. Let $r^\prime \leq r$, and let $\TT$ be a featured rooted tree of height at most $r^\prime$. We need to prove that there are distinct vertices $v_1,\dots, v_{k-1}$ in $\GG$ 
satisfying $\ball_{r^\prime}(v_i)\sim_{k-1,\epsilon,\eta} \TT$ that are
at distance at least $2r+1$ from each other, and at distance at least $2r+1$ from any cycle of length at most $2r+1$, from any root, and from $u$.
By assumption, $\GG$ is $(k,\epsilon,C, \frac{3r-1}{2})$-rich, so there are distinct
vertices $u_1,\dots, u_k$ satisfying $\ball_{r^\prime}(u_i)\sim_{k,\epsilon,\eta} \TT$ that are
at distance at least $6r+3$ from each other, and at distance at least $6r+3$ from any cycle of length at most $6r+3$, and from any root. In particular, $\ball_{r^\prime}(u_i)\sim_{k-1,\epsilon,\eta} \TT$ for all $i=1,\dots, k$. Observe that there is at most one vertex among $u_1,\dots, u_k$ at distance 
$2r+1$ from $u$. The remaining $k-1$ vertices can be chosen as $v_1,\dots, v_{k-1}$. This proves the result. 
\end{proof}
\section{Tools for Confirming that the Axioms Hold A.A.S. in the Linear Sparse Case} \label{app:linearsparseaastools}

In this section, we present tools for probability that we will use in showing that the axioms holds on most linear sparse graphs.

\subsection{Chaining Binomials}

We need a fact about chaining binomials:

\begin{lemma}
 \label{lem:aux_binom}
 Let $(X_n)_{n\geq 1}$ be a sequence of non-negative integer random variables satisfying
 $X_n/n \xrightarrow{p} \alpha$ for some
 $\alpha>0$. Let $(Y_n)_{n\geq 1}$ be another sequence of random
 variables over the same space satisfying
 $(Y_n \vert X_n = m) \sim \Bin(m,\beta)$ for all non-negative integers $m$ and
 some fixed $\beta\in [0,1]$. Then
 $Y_n/n \xrightarrow{p} \alpha \beta$.
\end{lemma}

\begin{proof}
    Let $0<\epsilon<\frac{2\alpha\beta}{1+\alpha}$. We need to show that:
    \[
    \lim_{n\to \infty}
    \Pr\left(
    \left\vert  \frac{Y_n}{n} - \alpha\beta
    \right\vert < \epsilon
     \right) = 1.
    \]
    Given an integer $m$, define $Y_{m,n}$ as
    $(Y_n \vert X_n = m)$.
    Let $\delta = \epsilon/(2\alpha\beta) < 1$. It holds that:
    \begin{align*}
    \Pr\left(
    \left\vert  \frac{Y_n}{n} - \alpha\beta
    \right\vert < \epsilon
     \right) \geq
    \sum_{1 - \delta \leq  m/\alpha n \leq 1+ \delta}
    \Pr(X_n=m) \Pr( \left\vert \Bin(m,\beta) - \alpha\beta n \right\vert < \epsilon n).
    \end{align*}
    By our choice of $\epsilon$ and $\delta$ we have:
    \begin{align*}
    \Pr( \left\vert \Bin(m,\beta) - \alpha\beta n \right\vert > \epsilon n) 
     &\leq 
    \Pr(\left\vert \Bin(m,\beta) - \beta m \right\vert > \epsilon n - | \alpha\beta n - \beta m |) \\ 
    &
    \leq 
    \Pr\left(\left\vert \Bin(m,\beta) - \beta m \right\vert > \frac{\epsilon}{2} n \right).    
    \end{align*}
    By our choice of $\epsilon$, it holds that $\frac{\epsilon n}{2 \beta m} < 1$.
    Using Chernoff's inequality for binomial variables we obtain that the last
    term is at most:
    \begin{align*}
    2 \exp{\left[ -
    \frac{1}{3}
    \left(\frac{\epsilon n}{2 \beta m}\right)^2
     \beta m    
    \right]}
      &\leq  
    2 \exp{\left[ -
    \frac{1}{3}
    \left(\frac{\epsilon n}{2 \beta (1+\delta) n}\right)^2
     \beta (1 - \delta) n
    \right]} \\ &
     \leq 
    2 \exp{\left[ -
    \frac{1}{3}
    \left(\frac{\epsilon^2(1 - \delta)}{4 \beta (1+\delta)^2}\right)
     n
    \right]}.
    \end{align*}
    This way, substituting in the first chain of inequalities we obtain
    that $\Pr(
    \vert  \frac{Y_n}{n} - $ $\alpha\beta
    \vert< \epsilon
     )$ at least:
    \begin{align*}
    & \Pr\left( \left\vert \frac{X_n}{n} - \alpha \right\vert \leq \alpha\delta \right) 
     \left(
     1- 2 \exp{\left[ -
    \frac{1}{3}
    \left(\frac{\epsilon^2(1 - \delta)}{4 \beta (1+\delta)^2}\right)
     n
    \right]}
     \right).
    \end{align*}
    The first probability tends to one with $n$ by assumption, and the second parentheses
    tends to one as well, so this completes the proof. 
\end{proof}

\subsection{Results about Branching Processes}
First, we need some information about branching processes.
We start with the expected size: 

\begin{fact}[Expected size of the branching process; {\cite[Theorem 3.3]{van_der_Hofstad_2024_Vol1}}]
\label{fact:size_BP}
    It holds that $\Ex\left[ \big\vert \BP\vert_r \big\vert \right] = c^r$. In particular, for every $\epsilon>0$, there is some $m$ such that
    $\Pr\left(
    \big\vert \BP\vert_r \big\vert > m
    \right) < \epsilon$.
\end{fact}


The notion of \emph{local convergence} \cite{van_der_Hofstad_2024_Vol2} expresses that the neighborhood distribution of a random graph sequence has a given limit distribution. For $\Gcal(n,c/n)$ this limit is given by the branching process
$\BP$, as stated in the following fact.

\begin{fact}[Local convergence to $\BP$; {\cite[Theorem 2.18]{van_der_Hofstad_2024_Vol2}}]
\label{fact:local_convergence_to_bp}
Let $r\geq 0$ be fixed, and let $\Tcal$ be a set of (non-featured) rooted trees of height at most $r$. Let $\BP\vert_r(\Tcal)$ be a shorthand for $\Pr(\BP\vert_r \in \Tcal)$.
Then for every $\epsilon>0$:
\[
\Pr_{\Gcal(n,c/n)}\left(
\left\vert
\frac{|\{ v\in [n] \, \vert \, \ball_r(v) \in \Tcal\}|}{n}
- \BP\vert_r(\Tcal) \right\vert > \epsilon
\right)
\]
tends to zero as $n$ goes to infinity.
\end{fact}

We prove the analogous result for the featured setting. 

\begin{theorem}[Local convergence to $\FBP$] \label{thm:local_convergence}
Let $r\geq 0$ be fixed, and let $\Tcal$ be a
(measurable) set of featured rooted trees of height at most $r$. 
 Let $\FBP\vert_r(\Tcal)$ be a shorthand for $\Pr(\FBP\vert_r \in \Tcal)$.
Then for every $\epsilon>0$:
\[
\Pr_{\Gcal_\Dcal(n,c/n)}\left(
\left\vert
\frac{|\{ v\in [n] \, \vert \, \ball_r(v) \in \Tcal\}|}{n}
- \FBP\vert_r(\Tcal) \right\vert > \epsilon
\right)
\]
tends to zero as $n$ goes to infinity.
\end{theorem}

\begin{proof}
Fix $\epsilon>0$. During this proof, identify $\BP$ with the underlying rooted tree
of $\FBP$. 
By \Cref{fact:size_BP}, there is a finite set of (non-featured) rooted trees $\mathcal{F}=$ $\{T_1,T_2,$ $\dots, T_\ell\}$ such that:
    \[
 \Pr(\BP\vert_r\in \mathcal{F}) > 1 - \epsilon/3.
    \]    
    For each $i=1,\dots, \ell$, let $X_{i,n}$ be the number of
    vertices $v$ in $\Gcal_{\Dcal}(n, c/n)$ 
    satisfying that the underlying 
    rooted graph of $\ball_r(v)$
    is isomorphic to $T_i$. Similarly, $Y_{i,n}$ is the variable counting the vertices
    $v$ in $\Gcal_{\Dcal}(n, c/n)$ for which this property above holds and additionally 
    $\ball_r(v)\in \Tcal$. We also define the constant $\rho_i$
    as:
    \[
    \Pr( \FBP\vert_r \in \Tcal \mid \BP\vert_r \sim T_i).
    \]
    By Fact \ref{fact:local_convergence_to_bp}, 
    $X_{i,n}/n \xrightarrow{p} \BP\vert_r(T_i)$. Also, as features are chosen independently from the underlying graph structure in $\Gcal_{\Dcal}(n,c/n)$,
    we have that
    $(Y_{i,n} \vert X_{i,n}= m) \sim \Bin(m, \rho_i)$. Hence, by Lemma \ref{lem:aux_binom}:
    \[
   \frac{Y_{i,n}}{n} \xlongrightarrow{p} 
    \rho_i\BP\vert_r(T_i)= \Pr(  \FBP\vert_r \in \Tcal \wedge \BP\vert_r \sim T_i
    ).
    \]
    Now we are ready to prove the statement. 
    Let $Z_n$ count the vertices $v$ in $\Gcal(n,c/n)$ satisfying that
    the underlying graph
     of $\ball_r(v)$ is not in $\mathcal{F}$,
    
    and let $W_n$ count how many such vertices also satisfy $\ball_r(v)\in \Tcal$.
    Let $\wt\Tcal$ be the subset of $\Tcal$ containing the featured trees whose underlying tree is not in $\mathcal{F}$.
    Consider the expression:
   \[ 
   \left\vert
\frac{|\{ v\in [n] \, \vert \, \ball_r(v) \in \Tcal\}|}{n}
- \FBP\vert_r(\Tcal) \right\vert
\]
    By the triangle inequality, this is at most: 
    \[
        \left\vert
        \frac{\sum_{i=1}^\ell Y_{i,n}}{n}
        - \sum_{i=1}^\ell \rho_i\BP\vert_r(T_i) \right\vert
        + \frac{W_n}{n} +  \FBP\vert_r(\wt\Tcal).
    \]
        We know that $W_n$ is bounded by above by $Z_n$,
        and $\FBP\vert_r(\wt\Tcal) \leq 
        1 - \BP\vert_r(\mathcal{F}) < \epsilon/3$,
        so the previous expression is at most:
        \[
\left\vert
\frac{\sum_{i=1}^\ell Y_{i,n}}{n}
- \sum_{i=1}^\ell \rho_i\BP\vert_r(T_i) \right\vert
+ \frac{Z_n}{n} +  \epsilon/3.
        \]
        Hence:
    \begin{gather*}
    \Pr\left(
\left\vert
\frac{|\{ v\in [n] \, \vert \, \ball_r(v) \in \Tcal\}|}{n}
- \FBP\vert_r(\Tcal) \right\vert > \epsilon
\right) 
\leq 
\Pr\left(
\left\vert
\frac{\sum_{i=1}^\ell Y_{i,n}}{n}
- \sum_{i=1}^\ell \rho_i\BP\vert_r(T_i) \right\vert  +  \frac{Z_n}{n} > \frac{2\epsilon}{3} \right).
    \end{gather*}
The last probability tends to zero as $n$ grows to infinity. Indeed, 
$\frac{Z_n}{n} \xrightarrow{p} c'$ for some $0< c' <  \frac{\epsilon}{3}$ 
by the definition of $\mathcal{F}$ and Fact \ref{fact:local_convergence_to_bp}, 
and $\frac{Y_{i,n}}{n} \xrightarrow{p} \rho_i\BP\vert_r(T_i)$ by Lemma \ref{lem:aux_binom}. This completes the proof.
\end{proof}

\subsection{Couplings}

We need  that appropriate couplings  exist for $X_0, X_1$ sufficiently different on a partition. 

\begin{lemma}[Constructing couplings] \label{lem:constructcoupling}
    Let $M$ be a set and $R\subseteq M^2$ a binary relation over $M$. Let $X_0,X_1$ be two random variables over $M$, and $\nu, \nu^\prime \geq 0$. Suppose $\{S_1,\dots, S_{\ell}\} \cup \{ T \}\subseteq 2^M$ is a partition of $M$, where $S_i \times S_i \subseteq R$ for each $i\in [\ell]$. Suppose that: 
    \[
    \abs*{\Pr(X_0\in S_i) - \Pr(X_1\in S_i)} \leq \frac{\nu}{\ell},
    \]
    for all $i\in [\ell]$, and:
    \[
    \Pr(X_0\in T) + \Pr(X_1\in T) \leq \nu^\prime.
    \]
    Then there is a coupling $\Pi$ of $X_0$ and $X_1$ satisfying:
    \[
    \Pr_{(s,t)\sim \Pi}( (s,t) \in R) \geq 1 - \nu - \nu^\prime.
    \]
\end{lemma}

\begin{proof}
    We construct the coupling $\Pi$ explicitly. We set $S_{\ell+1}=T$ for convenience.
    For each $i\in [\ell]$, $j=0,1$ we define $q_i= \min_{j=0,1} \Pr(X_j \in S_i)$, $p_{j,i}=  \max(0, \Pr(X_j\in S_i) - \Pr(X_{1-j} \in S_i))$. Finally, we define $q_{\ell+1}=0$, and $p_{j,\ell+1}= \Pr(X_j\in S_{\ell+1})$ for each $j=0,1$.
    Then
    it holds that:
    \begin{equation}
        \label{eq:coupling_aux0}
        \Pr(X_j \in S_i) = q_i + p_{j,i} \quad \text{for each $i\in [\ell+1], j=0,1$.}
    \end{equation}
    Define $p= 1 - \sum_{i\in \ell} q_i$. 
    Observe that:
    \[
    p= \sum_{i\in \ell} \abs*{\Pr(X_0\in S_i) - \Pr(X_1\in S_i)}
    + \Pr(X_0\in S_{\ell+1}) + \Pr(X_1\in S_{\ell+1}), 
    \] so by hypothesis it must be that:
    \begin{equation}
    \label{eq:couplig_aux2}
    p \leq \nu+ \nu^\prime
    \end{equation}
    We claim that:
     \begin{equation}
    \label{eq:coupling_aux1}
        \sum_{i\in [\ell+1]} p_{j,i} = p,
    \end{equation} 
    for $j=0,1$. Indeed, by \eqref{eq:coupling_aux0}:
    \[
    \sum_{i\in [\ell+1]} p_{0,i} - p_{1,i} = \sum_{i\in \ell} \Pr(X_0 \in S_i) - \Pr(X_1 \in S_i)= 0.
    \]
    So, in order to prove \eqref{eq:coupling_aux1} we just need to show that:
    \[
    \sum_{i\in [\ell+1]} p_{0,i} + p_{1,i} = 2p.
    \]
    For each $i\in [\ell+1]$ it holds that $\Pr(X_0\in S_i)+\Pr(X_1\in S_i)= 2q_i + p_{0,i} + p_{1,i}$, so we obtain:
    \[
    \sum_{i\in [\ell+1]} 2q_i + p_{0,i} + p_{1,i} = 2,
    \]
    and: 
    \[
    \sum_{i\in [\ell+1]} p_{0,i} + p_{1,i} = 2 - \sum_{i\in [\ell]} 2q_i = 2p,
    \]
    as we wanted to show. This proves \eqref{eq:coupling_aux1}. 
    We define an auxiliary random variable $W$.
    The range of $W$ is the set:
    \[\{ A_i \vert i\in [\ell]\} \cup \{B_i \vert i\in [\ell+1]\}^2.
    \]
    The role of the variable $W$ is to indicate whether $X_0$ and $X_1$ lie in a shared set $S_i$ for $i\in [\ell]$. In our desired coupling
    $W= A_i$ will mean that $X_0, X_1\in S_i$, while $W=(B_{i_0}, B_{i_1})$
    will mean that $X_0\in S_{i_0}$ and $X_1\in S_{i_1}$.
    We define $\Pr(W=A_i) = q_i$, and $\Pr(W= (B_{i_0}, B_{i_1}))= \frac{p_{0,i_0} p_{1,i_1}}{p}$ for each $i,i_0, i_1\in [\ell+1]$.
    In order to prove $W$ is a well-defined variable, we just need to show that the probabilities defining $W$ add up to one. That is:
    \begin{align*}
        \sum_{i\in [\ell+1]} \Pr(W=A_i) + 
        \sum_{i_0,i_1\in [\ell+1]} \Pr(W=(B_{i_0},B_{i_1}))
        &= 1 - p +
        \sum_{i_0\in [\ell+1]} p_{0,i_0} \left(
        \sum_{i_1\in [\ell+1]} \frac{p_{1,i_1}}{p}
        \right) \\ 
        &= 1 - p +
        \sum_{i_0\in [\ell+1]} p_{0,i_0}\cdot 1 = 1 - p + p = 1.
    \end{align*}
    We move on to constructing the coupling $\Pi$ of $X_0$ and $X_1$. For this we define a random vector $(\Pi_0, \Pi_1, \Pi_W)$. Then $\Pi$ will be defined as $(\Pi_0,\Pi_1)$. We have $\Pi_W\sim W$. For any $i\in [\ell+1]$ such that
    $q_i\neq 0$, the conditioned variables $\Pi_j\vert \Pi_W=A_i$ are independent for $j=0,1$,
    and $(\Pi_j\vert \Pi_W=A_i) \sim  (X_j \vert X_j \in S_i)$.  
    Similarly, let $i_0,i_1$ be such that $p_{0,i_0}, p_{1,i_1}\neq 0$.
    Then:
    \[(\Pi_j\vert \Pi_W=(B_{i_0},B_{i_1}))\sim
    (X_j \vert X_j\in S_{i_j})
    \]
    for each $j=0,1$ independently. Let us see that $(\Pi_0, \Pi_1)$ is a coupling of $X_0$ and $X_1$. We just need to show that
    $\Pi_j \sim X_j$ for each $j=0,1$. We prove the statement for $j=0$, the case $j=1$ is analogous. 
    The identity $\Pi_0 \sim X_0$ follows from the fact that for each $i\in [\ell+1]$ both (1)
    $(\Pi_0 \vert \Pi_0\in S_i) \sim (X_0 \vert X_0 \in S_i)$, and (2):
    \begin{align*}
    \Pr(\Pi_0 \in S_i) 
    &= \Pr(\Pi_W= A_i) + \sum_{i^\prime \in [\ell+1]}
    \Pr(\Pi_W = (B_i, B_{i^\prime})) \\  
    &= q_i + p_{0,i}\left(
    \sum_{i^\prime \in [\ell+1]} \frac{p_{1,i^\prime}}{p}
    \right) \\
    &= \Pr(X_0 \in S_i).        
    \end{align*}
    Here the last equality uses both \eqref{eq:coupling_aux0} and \eqref{eq:coupling_aux1}.  \par
    Now that we have constructed the coupling $\Pi=(\Pi_0, \Pi_1)$, all that is left is to show it satisfies the lemma's statement. Observe that when $\Pi_W=A_i$ for some $i\in [\ell]$, then $\Pi_0, \Pi_1\in S_i$, and
    $(\Pi_0, \Pi_1)\in R$. This way:
    \begin{align*}
    \Pr( (\Pi_0, \Pi_1) \notin R) 
    &\leq 1 - \sum_{i\in [\ell]} \Pr(\Pi_W=A_i) \\
    &= 1 - \sum_{i\in [\ell]} q_i = p \leq \nu+ \nu^\prime.
    \end{align*}
      Here the last inequality uses \eqref{eq:couplig_aux2}. This completes the proof.
\end{proof} 

\subsection{Valid Partitions}
We call a partition $\Pcal$ of $\FeatSp$ \emph{valid} if
it is measurable and $\Pr(\Dcal \in S)>0$ for each $S\in \Pcal$. 
Given a partition $\Pcal$ of $\FeatSp$ and a MRFG $\GG=(G, \bar v, \chi)$, the set $\Pcal(\GG)$ consists of all the MRFGs $\HH= (H, \bar u, \xi)$ satisfying that $|\bar v|=|\bar u|$ and there is an isomorphism $f: H  \rightarrow G$ that sends the $i^{th}$ root of $\HH$ to the $i^{th}$ root of $\GG$ for each $i\in [|\bar v|]$, and satisfies that $\xi(w)$ and $\chi(f(w))$ belong to the same set in $\Pcal$ for each $w\in V(H)$. In other words, $\HH$ is isomorphic to $\GG$ if we identify features according to $\Pcal$. \par

The following result states that when $\Pcal$ is a valid partition, and $\TT$ is a featured rooted tree, then the probability of obtaining $\TT$ as an outcome of $\FBP$ is positive, up to identifying features according to $\Pcal$.

\begin{lemma}[Featured trees modulo valid partitions have positive probabilities in $\FBP$]
    \label{lem:FBP_positive_probability}
    Let $\Pcal$ be a valid partition of $\FeatSp$, $r\geq 0$ an integer, and $\TT$ a featured rooted tree of height at most $r$. Then:
    \[
    \Pr( \FBP\vert_r \in \Pcal(\TT)) > 0.
    \]
\end{lemma}

\begin{proof}
    Let $\TT=(T,v, \chi)$. For each $u\in V(T)$, let $P_u\in \Pcal$ be the set containing $\chi(u)$. The features in $\FBP\vert_r$ are chosen independently from the underlying rooted tree. Hence:
    \[
         \Pr(\FBP\vert_r \simeq \TT) \geq \Pr(\BP\vert_r \simeq (T,v)) \prod_{u\in V(T)} \Pr(\Dcal \in P_u).
    \]
    To show the result it suffices to see that the right hand side of this inequality is bigger than zero. The fact that $\Pcal$ is a valid partition implies that $\prod_{u\in V(T)} \Pr(\Dcal \in P_u)$. Additionally, the probability that $\BP\vert_r\simeq (T,v)$ is at least: 
     \[
         \Pr(\Po_c = \deg(v)) \prod_{u\in V(T), u\neq v} \Pr(\Po_c = \deg(u)-1)> 0. 
     \] 
    Recall that $\Po_c$ denotes a Poisson variable with mean $c$. This completes the proof. 
\end{proof}

\subsection{Lifting Partitions}
The following lemma says that we can lift a partition of the feature space to a partition on MRFGs that plays nicely with the relations $\sim_{k,\epsilon,\eta}$.
\begin{lemma}[Partition lifting] \label{lem:partitionlifting}
    Let $k, m\geq 0$ be an integer, $\epsilon, C > 0$, and 
    $\Pcal$ a finite partition of $\FeatSp$ satisfying that $\lVert s - t \lVert_{\infty} \leq \epsilon$ for each $s,t\in P$, $P \in \Pcal$.
    Let $\Omega_{m}$ be the set of 
    MRFGs with $m$ roots.
    Then there is a partition $S_1,\dots, S_\ell$ of $\Omega_{r}$ satisfying:
    \begin{itemize}
        \item $\GG \sim^\dist_{k,\epsilon,\eta} \HH$ for each $\GG, \HH\in S_i$, $i\in [\ell]$.
        \item If $\GG\in S_i$ for some $i\in [\ell]$, then $\Pcal(\GG)\subseteq S_i$. 
    \end{itemize}
\end{lemma}

\begin{proof}
Fix $\epsilon, \eta, \Pcal$ as in the statement. We prove by induction on $k$
that the statement holds for all pairs $k,m\geq 0$. We define equivalence relations $\cong_{k,m}$ over $\Omega_{m}$ for each $k,m\geq 0$ that refine $\sim_{k,\epsilon,\eta}$ and have finite index. Then our desired partition of $\Omega_{m}$ will be the set of $\cong_{k,m}$-classes. \par
Let $k = 0$ and $m\geq 0$. Let $\GG=(G, \bar v, \chi), \HH=(H, \bar u, \xi)$ be two MRFGs in $\Omega_{m}$. We write 
$\GG \cong_{0,m} \HH$ for $\GG,\HH\in \Omega_{m}$
if the map matching the roots of $\GG$ to the roots of $\HH$ is a partial isomorphism between $G$ and $H$, and for each root $v$ of $\GG$ the corresponding root $u$ of $\HH$ satisfies that
$\chi(v)$ and $\xi(u)$ lie in the same set $P\in \Pcal$. Clearly the equivalence $\cong_{0,m}$ has a finite number of classes: the
number of non-isomorphic graphs of size at most $m$ is finite, and
the partition $\Pcal$ is finite as well. 
Moreover, $\cong_{0,m}$ refines $\sim^\dist_{0,\epsilon}$, and 
$\GG \cong_{0,m} \HH$ holds
for each $\GG\in \Omega_{m}$ and each $\HH\in \Pcal(\GG)$. Hence, the set of $\cong_{0,m}$-classes satisfies the statement. \par
Now let $k>0$ and $m\geq 0$ be arbitrary, and assume the statement holds for $k-1$ and all $m$. Let $S_1,\dots, S_\ell$ be a partition of $\Omega_{m+1}$
that witnesses the lemma for $k-1$ and $m+1$. Consider a finite partition $\Qcal$ of the interval $[0,1]$ satisfying $|s-t| \leq \frac{\eta}{\ell}$ for each $s,t\in Q$, $Q\in \Qcal$. Let $\GG,\HH\in \Omega_{m}$. We write
$\GG\cong_{k,m} \HH$ if (I) for each $v\in V(G)$ there is some $u\in V(H)$
such that $\GG[v]$ and $\HH[u]$ are in the same set $S_i$, and or each
$u\in V(H)$ there is some $v\in V(G)$ such that $\GG[v]$ and $\HH[u]$ are in the same set $S_i$, and (II) for each $i\in [m]$, the $i^{th}$ root $v_i$ of $\GG$ is isolated if and only if the $i^{th}$ root $u_i$ of $\HH$ is isolated as well. Otherwise, we require that for each $j\in [\ell]$ the numbers:
\[
\frac{\abs*{\{
v\in \Ne(v_i) \vert \GG[v] \in S_j
\}}}{|\Ne(v_i)|},
\]
and:
\[
\frac{\abs*{\{
v\in \Ne(v_i) \vert \GG[v] \in S_j
\}}}{|\Ne(v_i)|},
\]
belong to the same set $Q\in \Qcal$. \par
All that is left is to show that the set of $\cong_{k,m}$-classes satisfies the requirements from the lemma. Firstly, observe that there are a finite number of $\cong_{k,m}$-classes over $\Omega_{m}$. Indeed, the $\cong_{k,m}$-class of  
$\GG\in \Omega_{m}$ depends only on the sets $S_i$ containing $\GG[u]$ for each $u\in V(G)$, and the
set $Q\in \Qcal$ that contains the proportion of vertices $v\in \Ne(u)$ that belong to each set $S_i$, for each root $u$. Second, 
$\GG \sim_{k,\epsilon,\eta} \HH$ implies that $\GG \cong_{k,m} \HH$ for each $\HH,\GG\in \Omega_{m}$. This is straightforward to verify by induction. The fact that $\HH$ and $\GG$ satisfy the back and forth property follows from (I), and the neighborhood coupling property follows from (II), by applying Lemma \ref{lem:constructcoupling} to the partition $\{ S_1, \dots, S_\ell\}$, $T=\emptyset$, $\nu=\eta$, $\nu^\prime = 0$, 
and the relation $\cong_{k-1,m+1}$. Finally, it holds that $\GG \cong_{k,m} \HH$ for each $\GG\in \Omega_{m}$ and each $\HH\in \Pcal(\GG)$. This also can be shown by induction. 
\end{proof}
\section{Richness Holds A.A.S.: Proof of Theorem \ref{thm:axioms hold aas} \ref{item:richness; thm:axioms hold aas}} \label{app:richnessholds}

Recall Theorem \ref{thm:axioms hold aas} \ref{item:richness; thm:axioms hold aas}:

\medskip

    Let $k,r\geq 0$ be integers, and $\eta, \epsilon>0$. Then 
    a.a.s.\@ $\Gcal_\Dcal(n,c/n)$ is $(k,\epsilon,\eta,r)$-rich.  

\medskip

\begin{proof}
    We loosely follow \cite[Theorem 4.9]{lynchlinearsparse}. Let $\Pcal$ be a finite valid partition of $\FeatSp$ satisfying that $\lVert s-t \rVert_\infty \leq \epsilon$ for all $s,t\in P$, $P\in \Pcal$. Such a partition exists because $\FeatSp$ is a compact set. 
    Let $\Omega_{1}$ be the set of MRFGs containing a single root, and let $\mathcal{S}$ be a finite partition of $\Omega_{1}$ obtained by applying  the partition lifting lemma, Lemma~\ref{lem:partitionlifting}, to $k,\epsilon, \eta,\Pcal$ and $m=1$. 
    Let $\mathcal{T}\subseteq \mathcal{S}$ be the  partitions $S \in \mathcal{S}$ containing some tree of height at most $r$. \par
    We introduce an auxiliary definition. Let $S\in \Tcal$. A MFRG $\GG$ is $(k,S)$-rich if
    there are distinct vertices $v_1,\dots, v_k \in V(G)$ 
    satisfying that 
    \begin{enumerate}[label=(\arabic*)]
        \item
        $\ball_{r_k}(v_i) \in S$        
        for all $i\in [k]$.
        \item For distinct $i,j\in [k]$ the distance between $v_i$ and $v_j$ is greater than $2r_k+1$.
        \item For all $i\in [k]$, the
        distance from $v_i$ to any root of $\GG$ or any cycle of length at most $2r_k +1 $ is greater than
        $2r_k + 1$.
    \end{enumerate}
    By our choice of $\Tcal$, if a MRFG $\GG$ is $(k,S)$-rich for each $S\in \Tcal$, then it is $(k,\epsilon, \eta, r)$-rich. We prove this stronger property holds a.a.s. in $\Gcal_{\Dcal}(n,c/n)$. In other words, 
    each $S\in \Tcal$, a.a.s. $\Gcal_{\Dcal}(n,c/n)$ is $(k,S)$-rich. Because there are only finitely many $S\in \Tcal$, this proves the result. \par
    Let $S\in \Tcal$. Fix $\nu >0$. We show that the probability that $\Gcal_{\Dcal}(n,c/n)$ is $(k,S)$-rich is bigger than $1-\nu$ for sufficiently large $n$. Define $\rho=
    \Pr(\FBP\vert_{r_k} \in S)$. Let $\TT\in S$ be a featured rooted tree. By the construction of $\Tcal$, it holds that $\Pcal(\TT)\subseteq S$. Recall now Lemma \ref{lem:FBP_positive_probability}, which states that each tree has a positive probability in $\FBP\vert_{r_k}$ when features are identified up to a valid partition. Applying the lemma, we see that:
    \[
    \Pr(\FBP\vert_{r_k} \in \Pcal(\TT)) > 0,
    \]
    so in particular $\rho > 0$. Fix an integer
    $m$ such that:
    \[
    \binom{m}{k} (1-\rho)^{m-k} < \nu /2.
    \]
    Such $m$ exists because the left-hand side of this inequality is asymptotically equivalent to
    $m^k (1-\rho)^m$ as $m$ grows to infinity, and $0 \leq 1-\rho< 1$.
    Let $B_n$ be the event that there are at least $k$ vertices $v$ among $1,\dots, m$ such that $N(v,r_k)\in S$ in $\Gcal_\Dcal(n,c/n)$. Using Theorem \ref{thm:local_convergence}, local convergence to $\FBP$, and the intersection bound, 
    we get that:
    \[
    \Pr(B_n) \geq 1 - \binom{m}{k} (1-\rho_i)^{m-k} + o(1) \geq 1 - \nu/2 + o(1).
    \]
    Let $A_n$ be the event that no two of the vertices $1,\dots, m$ in $\Gcal_\Dcal(n,c/n)$ are at distance less or equal than $2r_k+1$, and none of these vertices is at distance less or equal than $2r_k + 1$
    to a cycle of size at most $2 \cdot r_k + 1$. By  a first moment argument  $\Pr(A_n)= 1 -
    O(1/n)$: given $r\leq 2r_k+1$, the expected number of paths of length $r$ whose endpoints are in $[m]$ is: \[
    O\left(m^2 n^{r-1} \left(\frac{c}{n}\right)^r \right) = O\left(\frac{1}{n}\right).\]
    Adding up for all $r\leq r_k$, we obtain that the expected number of pairs $u,v\in [m]$ that are at distance at most $2r_k+1$ is $O(1/n)$, so the probability that such a pair exists is $O(1/n)$ by Markov's inequality. A similar argument works for bounding the expected number of vertices $v\in [m]$ that are within distance $2r_k+1$ to some cycle of length at most $2r_k + 1$ is also $O(1/n)$. This shows that $\Pr(A_n)= 1 - O(1/n)$. Using the intersection bound again, we get that:
    \[
    \Pr(B_n \wedge A_n) \geq 1 - \nu/2 + O(1/n).
    \]
    Observe that when both $B_n$ and $A_n$ hold, then 
     $\Gcal_{\Dcal}(n,c/n)$ is $(k,S)$-rich. As our choice of $\nu$ was arbitrary, this
    property holds a.a.s., completing the proof.   
\end{proof}
\section{FBP Axioms Hold A.A.S.: Proof of Theorem \ref{thm:axioms hold aas} \ref{item:fpb; thm:axioms hold aas}} \label{app:bpholds}

Recall Theorem \ref{thm:axioms hold aas} \ref{item:fpb; thm:axioms hold aas}:

\medskip

    Let $k,r\geq 0$ be integers, and $\eta, \epsilon>0$. Then 
    a.a.s. $\Gcal_\Dcal(n,c/n)$ is $(k,\epsilon, \eta, r)$-similar to FBP.

\medskip

\begin{proof}
Let $\Pcal$ be a finite valid partition of $\FeatSp$ satisfying that $\lVert s-t \rVert_\infty \leq \epsilon$ for all $s,t\in P$, $P\in \Pcal$. Such a partition exists because $\FeatSp$ is compact. Let $\Omega_1$ be the set of MRFGs containing a single root. Let $\mathcal{S}$ be a finite partition of $\Omega_{1}$ obtained by applying the Partition Lifting lemma, Lemma~\ref{lem:partitionlifting}, to $k, \epsilon, \eta, \Pcal$ and $m=1$.  
Let $\{S_1,\dots, S_\ell\} \subseteq \mathcal{S}$ be the set of partition elements $S_i \in \mathcal{S}$ containing some tree of height at most $r$.  Using local convergence to $\FBP$, Theorem \ref{thm:local_convergence}, we obtain that a.a.s.:
    \begin{align*}
        \sum_{i \in [\ell]} 
        \left|
            \frac{|\{ v\in [n] \vert \Ne^{\Gcal_\Dcal(n,c/n)}_{r}(v) \in S_i \}|}{n} - \Pr(\FBP\vert_{r} \in S_i)
        \right| \leq \frac{\eta}{2\ell}. 
    \end{align*}
    Let $T$ denote the set of MRFGs are not contained in some class $S_i$. In particular, $T$ contains no featured rooted tree of height at most $r$.
    Clearly $\Pr(\FBP\vert_{r}\in T)=0$, so using the local convergence theorem to BP, Theorem \ref{thm:local_convergence}, we obtain that a.a.s.:
    \[
       \sum_{i \in [\ell]} 
        \left|
            \frac{|\{ v\in [n] \vert \Ne^{\Gcal_\Dcal(n,c/n)}_{r}(v) \in T \}|}{n}
        \right| \leq \frac{\eta}{2}.
    \]
    Let $\HH_n$ be the random MRFG obtained by considering $\ball_{r}(v)$ in $\Gcal_\Dcal(n,c/n)$, where $v\in [n]$ is chosen uniformly at random. 
    The inequalities above shows we can use the Constructing Couplings lemma,
    Lemma \ref{lem:constructcoupling},
    on the random variables $\HH_n$, $\FBP\vert_r$, the binary relation $\sim_{k,\epsilon,\eta}$, the partition $\{ S_1,\dots, S_\ell\} \cup \{T\}$, and the constants $\nu=\nu^\prime = \frac{\eta}{2}$.
    Hence, a.a.s. there is a coupling $\Pi$ of $\HH_n$ and $\FBP\vert_{r}$ satisfying that:
    \[
    \Pr_{(\HH, \TT)\sim \Pi}(\HH \sim_{k,\epsilon,\eta} \TT) \geq  1 - \eta.
    \]
    This proves the result.
\end{proof}
\section{Homogeneity Holds A.A.S.: Proof of Theorem \ref{thm:axioms hold aas} \ref{item:homogeneity; thm:axioms hold aas}} \label{app:homogeneity}

Recall Theorem \ref{thm:axioms hold aas} \ref{item:homogeneity; thm:axioms hold aas}:

\medskip
    Let $k,r\geq 0$ be integers, and $\eta>0$. Then a.a.s.\@ $\Gcal_\Dcal(n,c/n)$ is $(k, \eta, r)$-homogeneous.
\medskip

We start with the following result.
\begin{lemma} 
\label{lem:aux_homogeneous}
    The following statements hold in $\Gcal_\Dcal(n,c/n)$:
    \begin{enumerate}[label=(\Roman*)]
        \item A.a.s.\@ $\Delta(\Gcal_\Dcal(n,c/n))$ is at most $\ln n$. 
        \item For any fixed $r\geq 0$, a.a.s.\@ the number of vertices in $\Gcal_\Dcal(n,c/n)$ belonging to cycles of length at most $r$ is at most $\ln n$.
    \end{enumerate}
\end{lemma}
\begin{proof}
    Both items are direct applications of the first moment method. We start with the first. 
    The probability that a given vertex has degree at least $\ln n$ in $\Gcal(n, c/n)$ is at least
    \[
    \binom{n}{\ln n} \left( \frac{c}{n} \right)^{\ln n} = O\left(
    \left(
    \frac{c}{e \ln n}
    \right)^{\ln n}
    \right) =  O\left(
    \frac{1}{n}
    \left(
    \frac{c}{\ln n}
    \right)^{\ln n}
    \right),
    \]
    where we have used Stirling's approximation for the binomial coefficient. Hence, the expected number of vertices with degree at least $\ln n$ is
    \[
     O\left(
    \left(
    \frac{c}{\ln n}
    \right)^{\ln n}
    \right),
    \]
    which tends to zero with as $n$ grows to infinity. \par
    To see the last item, observe that the expected number of $\ell$-cycles in $\Gcal(n,c/n)$
    is $\frac{c^\ell}{2\ell}+ o(1)$. Hence, the expected number of vertices in cycles of size at most $r$
    is
    \[
    \sum_{i=3}^r i \frac{c^i}{2i} + o(1) \leq \frac{1}{2} \sum_{i=1}^r c^i + o(1).
    \]
    Let $\nu=  \frac{1}{2} \sum_{i=1}^r c^i$. Then by Markov's inequality the probability that there are at least $\ln n$ vertices lying in cycles of length at most $r$ is 
    $\frac{\nu}{\ln n} + o(1)$, which tends to zero. This completes the proof.      
\end{proof}

Theorem \ref{thm:axioms hold aas} \ref{item:homogeneity; thm:axioms hold aas} follows from Lemma \ref{lem:aux_homogeneous} by unrolling the definition of homogeneity.

\end{document}